\documentclass{aastex6}
\usepackage[latin9]{inputenc}
\usepackage{array}
\usepackage{amsbsy}
\usepackage{graphicx}

\makeatletter

\providecommand{\tabularnewline}{\\}

\makeatother

\begin{document}

\title{On the Origin of Solar Torsional Oscillations and Extended Solar
Cycle}

\author{V.V. Pipin$^{1}$, A.G. Kosovichev$^{2,3}$}

\affil{$^{1}$Institute of Solar-Terrestrial Physics, Russian Academy of
Sciences, Irkutsk, 664033, Russia\\
 $^{2}$Department of Physics, New Jersey Institute of Technology,
Newark, NJ 07102, USA\\
 $^{3}$Center for Computational Heliophysics, New Jersey Institute
of Technology, Newark, NJ 07102, USA}
\begin{abstract}
We present a nonlinear mean-field model of the solar interior dynamics
and dynamo, which reproduces the observed cyclic variations of the
global magnetic field of the Sun, as well as the differential rotation
and meridional circulation. Using this model, we explain, for the
first time, the extended 22-year pattern of the solar torsional oscillations,
observed as propagation of zonal variations of the angular velocity
from high latitudes to the equator during the time equal to the full
dynamo cycle. In the literature, this effect is usually attributed
to the so-called ``extended solar cycle''. In agreement with the
commonly accepted idea our model shows that the torsional oscillations
can be driven by a combinations of magnetic field effects acting on
turbulent angular momentum transport, and the large-scale Lorentz
force. We find that the 22-year pattern of the torsional oscillations
can result from a combined effect of an overlap of subsequent magnetic
cycles and magnetic quenching of the convective heat transport. The
latter effect results in cyclic variations of the meridional circulation
in the sunspot formation zone, in agreement with helioseismology results.
The variations of the meridional circulation together with other
drivers of the torsional oscillations maintain their migration to
the equator during the 22-year magnetic cycle, resulting in the observed
extended pattern of the torsional oscillations.  
\end{abstract}

\section{Introduction}

It is widely accepted that differential rotation is one of the most
important causes of solar magnetic activity. In the kinematic hydromagnetic
dynamo regime, the differential rotation generates large-scale toroidal
magnetic field by stretching the poloidal magnetic field component,
and the poloidal magnetic field is generated by helical turbulent
convection from the toroidal component \citep{Parker1955}. In the
nonlinear dynamical regime, the dynamo-generated magnetic fields can
affect the differential rotation as well as turbulent dynamo processes.
Thus, observational evidences of the interaction of magnetic field
and differential rotation can provide important constraints on theoretical
models of the solar dynamo. Since the discovery of zonal variations
of the angular velocity (``torsional oscillations'') by \citet{Labonte1982}
it was found that these variations represent a complicated wave-like
pattern which consists of alternating zones of accelerated and decelerated
plasma flows \citep{Snodgrass1985,Altrock2008,Howe2011}. \citet{Ulrich2001}
found that the wave pattern consists of two oscillatory modes with
the periods of 11 and 22 years. Torsional oscillations were linked
to ephemeral active regions that emerge at high latitudes during the
declining phase of solar cycles, but represent magnetic field of the
following cycle \citep{Wilson1988}. In addition, the torsional oscillations
were linked to the migrating pattern of coronal green-line emission.
These observational results led to the concept of a 22-year long ``extended
solar cycle''\citep{Altrock1997}. Using results of the solar magnetic
field observations, \citet{Stenflo1988} and \citet{Sten1992} showed
that superposition of different eigenmodes of the global dynamo can
result in large-scale magnetic field patterns with extended and overlapping
branches in time-latitude diagrams. Similar patterns were reproduced
by the mean-field dynamo models, e.g., \citet{Brandenburg1991} or
\citet{Pipin2012}. Direct numerical simulation are capable to reproduce
them as well, e.g., \citet{Kapyla2016}.\textbf{ }Doppler measurements
of the solar rotation by \citet{Ulrich2001} as well as results of
helioseismic inversion by \citet{Howe2018} and \citet{Kosovichev2019}
have demonstrated properties of the ``extended'' 22-year variations
of the zonal flows. In these variations, the zonal flow pattern drifts
poleward and equatorward, starting at about 55-degree latitude nearly
simultaneously with activity of solar ephemeral active regions of
a new solar cycle. The extended mode travels to the equator with the
period equal to the full solar magnetic cycle, i.e., about 22 year.
The polar-ward branch reaches the polar regions by the time of the
solar maximum.

Analysis of helioseismology data for almost two solar
cycles by \citet{Kosovichev2019} revealed zones of deceleration of
the torsional oscillations, which at the surface corresponds to regions
of emerging magnetic field (Fig.~\ref{figure01}). The torsional
oscillation pattern in the near-surface layers obtained by subtracting
the mean differential rotation {(separately for Solar Cycles 23 and
24)}, and combining the residuals in the time-latitude diagram is
shown in Fig.~\ref{figure01}b. For comparison with the evolution
of solar magnetic field, in Fig.~\ref{fig1}a we present the magnetic
butterfly diagram showing the evolution of radial magnetic field.
Comparison with the zonal velocity diagram shows that the active regions
predominantly emerge at the boundary between the fast and slow zones,
at which the fast zone is closer to the equator than the slow zone,
as was found in the previous studies. The corresponding zonal acceleration
which reflects forcing of the torsional oscillations at the solar
surface is shown in Fig.~\ref{figure01}c. It clearly reveals the
torsional oscillation patterns of both Cycles 23 and 24. By overlaying
the zonal acceleration and magnetic field diagrams (Fig.~\ref{figure01}d)
we find that the active region zones coincide with the flow deceleration
zone (blue color). In the polar regions, the deceleration zones correspond
to the periods of strong polar magnetic field. This confirms the original
ideas that the torsional oscillations are due to the back reaction
of solar magnetic fields. Detailed analysis showed that the zonal
deceleration originates near the bottom of the convection zone at
high latitudes, and migrates to the surface corresponding to magnetic
dynamo waves predicted by the Parker's dynamo theory. On the surface,
the torsional oscillation pattern displays two branches migrating
poleward and equatorward. The torsional oscillation cycle takes about
22 years. It is important to note that if the torsional oscillation
pattern is mainly mediated by the Lorentz force of the near-surface
magnetic field, it has to change the sign twice during the full 22-year
magnetic cycle. However, in this case, the extended 22-year appearance
is not explained. This means that the mechanism of the extended solar
cycle phenomenon is related to processes in the deep solar interior.
Initially, \citet{Malkus1975} showed that the Lorentz force affects
efficiency of the dynamo process, and \citet{Yoshimura1981} suggested
the first non-kinematic dynamo model that explained the zonal flow
variations with the 11-year period. Beside the Lorentz force, other
suggested mechanisms of the solar torsional oscillations are related
to magnetic perturbation of the Taylor-Proudman balance, and effects
of magnetic fields on the heat transport in the solar convection zone
\citep{Spruit2003,Rempel2006}. Other mean-field dynamo models \citep[e.g.][]{Kueker1996,Pipin2018b}
and 3D MHD numerical simulations \citep[e.g.][]{Beaudoin2013,Guerrero2016a,Kapyla2016}
also showed that 11-year zonal variations of differential rotation
can be caused by the Lorentz force. The simulations produced variations
resembling the observed of torsional oscillations, potentially including
the extended mode. However, these models did not correctly reproduce
the observed cycle duration and time-latitude diagrams of the torsional
oscillations, as well as the observed phase relation between the surface
magnetic and flow fields (Fig.~\ref{figure01}). 

In this paper, we present a non-linear mean-field model of the solar
dynamo coupled with large-scale dynamics, which explains the extended
pattern of the torsional oscillations as well as other properties
observed on the solar surface and in the convection zone. The model
reproduces well-known properties of the solar dynamo waves that show
the \textbf{22-year cyclic} mode \citep{Stenflo1988}. We extend analysis
of the nonlinear dynamo model \citep{Pipin2018b} that includes effects
of the dynamo-generated magnetic field on the angular momentum and
heat transport in the solar convection zone. In particular, we discuss
the non-dissipative angular momentum flux that is generated by the
large-scale magnetic field. In the mean-field hydrodynamics, the non-dissipative
angular momentum fluxes (called the $\Lambda$-effect) are parametrized
by a tensor \citep{Ruediger1989}. Theoretically, it was predicted
that the large-scale magnetic field produces additional components
of the $\Lambda$-effect, which can induce the differential rotation
\citep{Kitchatinov1994,Kueker1996}. This effect was confirmed by
numerical simulations of \citet{Kapyla2019}.

Our model utilizes the concept of mean-field turbulent dynamo operating
in the bulk of the solar convection zone. Some arguments for this
type of dynamo were discussed by \citet{Brandenburg2005,Pipin2011a}
and \citet{Kosovichev2013}. In this model, the observed butterfly
diagram is explained by migrating dynamo waves, the diffusive migration
of which is modified by effects of turbulent pumping and meridional
circulation. The effective drift of the large-scale magnetic field
because of pumping and meridional circulation was discussed by \citet{Pipin2018b}.
There is still controversy about which type of dynamo operates on the solar convection zone. Therefore, we formulate our model in such a way that, in addition to the dynamo process, it describes the background large-scale hydrodynamic flows, and thus, can be used for the case of the flux-transport dynamo, as well. Among the main conditions for the flux-transport dynamo are the presence of global one-cell meridional circulation and tachocline at the bottom of the convection zone \citep{Dikpati1999,Jouve2008}. The impact of these conditions on the solar/stellar dynamo is a highly debated topic (see, e.g., \citealt{Zhao2014,Boning2017,Rajaguru2015,Wright2016,Cameron17}).

The model presented in this paper belongs to the class of distributed
dynamos, and, depending on model parameters, can describe both cases
with the single- and double-cell meridional circulation. This dynamical
mean-field model reproduces the cyclic behavior of the large-scale
solar magnetic field, as well as the differential rotation, meridional
circulation and their variations with the solar cycle. We analyze
effects of various forcing terms (by turning them on and off) on solar-cycle
variations of the differential rotation, and show that the observed
extended pattern of the solar torsional oscillations results from
the overlap of the 22-year dynamo cycles and effects of dynamo-generated
magnetic fields on the turbulent angular momentum and heat transport
in the solar convection zone.

\section{Basic equations.}

\subsection{\label{subsec:The-heat-transport}The heat transport and angular
momentum balance}

We employ the standard evolutionary model of the solar thermodynamic
structure, which is calculated using the MESA code \citep{Paxton2011,Paxton2013}.
The convective turnover time $\tau_{c}$, determined from the MESA
code as a function of radius, is assumed to be independent of time.
The RMS of convective velocity, $\mathrm{\mathrm{u}_{c}}$, is calculated
in the mixing-length approximations from the gradient of mean entropy,
$\overline{\mathrm{s}}$, 
\[
\mathrm{u_{c}=\frac{\ell_{c}}{2}\sqrt{-\frac{g}{2c_{p}}\frac{\partial\overline{s}}{\partial r}},}
\]
where $\ell_{c}=\alpha_{MLT}H_{p}$ is the mixing length,\textbf{
$\alpha_{MLT}=1.86$ }is the mixing-length theory parameter, and $H_{p}$
is the pressure scale height. The mean-field equation for heat transport
is described by the energy conservation law, and takes into account
effects of rotation and magnetic field \citep{Pipin2000}: 
\begin{equation}
\overline{\rho}\overline{T}\left(\frac{\partial\overline{\mathrm{s}}}{\partial t}+\left(\overline{\mathbf{U}}\cdot\boldsymbol{\nabla}\right)\overline{\mathrm{s}}\right)=-\boldsymbol{\nabla}\cdot\left(\mathbf{F}^{c}+\mathbf{F}^{r}\right)-\hat{T}_{ij}\frac{\partial\overline{U}_{i}}{\partial r_{j}}-\boldsymbol{\boldsymbol{\mathcal{E}}}\cdot\left(\nabla\times\overline{\boldsymbol{B}}\right),\label{eq:heat}
\end{equation}
where $\hat{\mathbf{T}}$ is the turbulent stress tensor that includes
small-scale fluctuations of velocity and magnetic field: 
\begin{equation}
\hat{T}_{\mathrm{ij}}=\overline{u_{i}u_{j}}-\frac{1}{4\pi\overline{\rho}}\left(\overline{b_{i}b_{j}}-\frac{1}{2}\delta_{ij}\overline{\mathbf{b}^{2}}\right),\label{eq:stres}
\end{equation}
$\mathbf{u}$ and $\mathbf{b}$ are the turbulent fluctuating velocity
and magnetic field, respectively. Other quantities in Eq.(\ref{eq:heat})
include the mean electromotive force, $\boldsymbol{\mathcal{E}}=\left\langle \mathbf{u\times b}\right\rangle $,
$\overline{\rho}$ is the mean density, $\overline{T}$ - the mean
temperature, $\mathbf{F}^{c}$ - the eddy convective flux, and $\mathbf{F}^{r}$
is the radiative flux. For calculation of $\hat{\mathbf{T}}$, $\boldsymbol{\boldsymbol{\mathcal{E}}}$
and $\mathbf{F}^{c}$, we employ analytical results that were obtained
earlier using the mean-field magnetohydrodynamics framework. These
results take into account effects of the global rotation and magnetic
field on turbulence. Some important details and references are given
below and in Appendix.

The magnitude of $\hat{\mathbf{T}}$, $\boldsymbol{\boldsymbol{\mathcal{E}}}$
and $\mathbf{F}^{c}$ in Eq.(\ref{eq:heat}) depends on the RMS of
convective velocity, $\mathrm{\mathrm{u}_{c}}$, efficiency of the
Coriolis force, and the strength of large-scale magnetic field. The
effect of the Coriolis force is determined by parameter $\Omega^{*}=2\Omega_{0}\tau_{c}$,
where $\tau_{c}$ is the convective turnover time. We assume that
the characteristic solar rotation rate corresponds to the rotation
rate of the tachocline at 30$^{\circ}$ latitude, i.e., $\Omega_{0}/2\pi=430$nHz
\citep{Kosovichev1997}. The influence of large-scale magnetic field
on convective turbulence is determined by parameter $\mathrm{\beta=\left|\overline{\mathbf{B}}\right|/\sqrt{4\pi\overline{\rho}u_{c}{}^{2}}}$.

For the anisotropic convective flux, we employ the expression suggested
by \citet{Kitchatinov1994a}: 
\begin{equation}
\mathrm{F_{i}^{c}=-\overline{\rho}\overline{T}\chi_{ij}\nabla_{j}\overline{s}.}\label{conv}
\end{equation}
Further details about dependence of the eddy thermal conductivity
tensor, $\mathrm{\chi_{ij}}$, on the global rotation and large-scale
magnetic field are given in Appendix A, in Eq.(\ref{eq:htfl}).

For the angular momentum balance, we use the model recently developed
by \citet{Pipin2018b}. The model describes evolution of the mean
axisymmetric velocity: $\mathrm{\mathbf{\overline{U}}=\mathbf{\overline{U}}^{m}+r\sin\theta\Omega\hat{\mathbf{\boldsymbol{\phi}}}}$,
where $\boldsymbol{\hat{\phi}}$ is the azimuthal unit vector, and
$\mathbf{\overline{U}}^{m}$ is the meridional circulation velocity.
We employ the anelastic approximation. Conservation of the angular
momentum determines distribution of the angular velocity inside the
convection zone: 
\begin{eqnarray}
\frac{\partial}{\partial t}\overline{\rho}r^{2}\sin^{2}\theta\Omega & = & -\boldsymbol{\nabla\cdot}\left(r\sin\theta\overline{\rho}\left(\hat{\mathbf{T}}_{\phi}+r\sin\theta\Omega\mathbf{\overline{U}^{m}}\right)\right)\label{eq:angm}\\
 & + & \boldsymbol{\nabla\cdot}\left(r\sin\theta\frac{\overline{\mathbf{B}}\overline{B}_{\phi}}{4\pi}\right).\nonumber 
\end{eqnarray}
The meridional circulation is determined from equation for the azimuthal
component of large-scale vorticity, $\mathrm{\overline{\omega}=\left(\boldsymbol{\nabla}\times\overline{\mathbf{U}}^{m}\right)_{\phi}}$:
\begin{eqnarray}
\mathrm{\frac{\partial\omega}{\partial t}\!\!\!} & \mathrm{\!\!=\!\!\!\!} & \mathrm{r\sin\theta\boldsymbol{\nabla}\cdot\left(\frac{\hat{\boldsymbol{\phi}}\times\boldsymbol{\nabla\cdot}\overline{\rho}\hat{\mathbf{T}}}{r\overline{\rho}\sin\theta}-\frac{\mathbf{\overline{U}}^{m}\overline{\omega}}{r\sin\theta}\right)}\label{eq:vort}\\
 & + & \mathrm{r}\sin\theta\frac{\partial\Omega^{2}}{\partial z}-\mathrm{\frac{g}{c_{p}r}\frac{\partial\overline{s}}{\partial\theta}}+F_{L}^{(p)}\nonumber 
\end{eqnarray}
where $\mathrm{\partial/\partial z=\cos\theta\partial/\partial r-\sin\theta/r\cdot\partial/\partial\theta}$
is the gradient along the axis of rotation. The first line in Eq.(\ref{eq:vort})
describes dissipation and advection of the large-scale vorticity and
meridional circulation; the second line describes effects of the centrifugal
force, the thermal wind, and the poloidal component of the large-scale
Lorentz force, $F_{L}^{(p)}$ (see, the Eq(\ref{flp})).

Both, the eddy thermal conductivity, $\chi_{T}$, and viscosity $\nu_{T}$,
are determined from the mixing-length theory: 
\begin{eqnarray*}
\chi_{T} & = & \frac{\ell^{2}}{6}\sqrt{-\frac{g}{2c_{p}}\frac{\partial\overline{s}}{\partial r}},\\
\nu_{T} & = & \mathrm{Pr}_{T}\chi_{T},
\end{eqnarray*}
where $\mathrm{Pr}_{T}$ is the turbulent Prandtl number. It is assumed
that $\mathrm{Pr}_{T}=3/4$.

The described model of the mean-field heat transport and angular momentum
balance follow the line of work of \citet{Kitchatinov1994}. In terms
of the original theory of the $\Lambda$-effect, the model reproduces
results of \citet{Kitchatinov2011a}. In this case, the mean-field
model reproduces the solar-like differential rotation profile with
one meridional circulation cell per hemisphere. The double-cell meridional
circulation structure is reproduced when inversion of the $\Lambda$-effect
radial profile in the solar convection zone is taken into account
\citep{Bekki2017}. Such inversion can result from radial inhomogeneity
of the convective turnover time, $\tau_{c}$ \citep{Pipin2018c}.
In our model, it is calculated using the mixing-length approximation,
$\tau_{c}=\ell_{c}/\mathrm{u_{c}}$. According to the convection zone
properties given by the MESA code, $\tau_{c}$ increases sharply towards
the bottom of convection zone. This results in a secondary meridional
circulation cell \citep{Pipin2018c}. The helioseismology results
are still contradictory about the strength and parameters of the meridional
circulation near the bottom of the convection zone (see, e.g., \citealt{Zhao2014,Boning2017,Rajaguru2015,Wright2016,Cameron17}).
Therefore, we consider both cases of the single- and double-cell meridional
circulation structure. The smooth transition between these two cases
can be controlled by variations of the mixing-length parameter, $\ell_{c}$,
using the following ansatz of \citet{Kitchatinov2017}: 
\begin{equation}
\ell_{c}=\ell_{\mathrm{min}}+\frac{1}{2}\left(\ell_{c}^{(0)}-\ell_{\mathrm{min}}\right)\left[1+\mathrm{erf}\left(\frac{r-(r_{b}+\ell_{\mathrm{min}})}{R_{\odot}d}\right)\right],\label{lmin}
\end{equation}
where $\ell_{c}^{(0)}$ is the mixing-length parameter from the MESA
code, $r_{b}=0.728R_{\odot}$ is the radius of the bottom of the convection
zone, $d=0.02$. We use $\ell_{\mathrm{min}}$ as a control parameter
to model saturation of $\tau_{c}$ variations in the $\Lambda$-tensor.
For $\ell_{\mathrm{min}}\le0.01R_{\odot}$, we obtain solutions with
the double cell meridional circulation structure. In another development
of the models of \citet{Pipin2018c} and \citet{Pipin2018b}, we add
the tachocline. Similarly to \citet{Rempel2006}, we use a phenomenological
approach to model the angular velocity profile in the tachocline.
Within the tachocline layer, we solve Eq.(\ref{eq:angm}) assuming
continuity of the stress and angular velocity at the bottom of the
convection zone and the solid body rotation state of the radiative
zone at inner boundary, $r_{t}=0.68R_{\odot}$.

The turbulent parameters below the convection zone are defined following
the results of \citet{Ludwig1999}, also see \citet{Paxton2011}.
We apply an exponential decrease of all turbulent coefficients (except
the eddy viscosity and eddy diffusivity) with decrement -100, i.e.,
they are multiplied by a factor of $\exp\left(-100z/R\right)$, where
$z$ is the distance from the bottom of the convection zone. We keep
the eddy viscosity and eddy diffusivity finite at the bottom of the
tachocline, i.e., for the eddy viscosity coefficient profile within
the tachocline we put 
\begin{equation}
\nu_{T}^{(t)}=\frac{\nu_{T}^{(c)}}{\left(\nu_{T}^{(0)}+\nu_{T}^{(c)}\right)}\left(\nu_{T}^{(0)}+\nu_{T}^{(c)}\exp\left(-100z\right)\right),\label{nut}
\end{equation}
where $\nu_{T}^{(c)}$ is the value at the bottom of the convection
zone, $\nu_{T}^{(0)}$ is the value inside the tachocline, $z$ is
the distance from the bottom of the convection zone. We use $\nu_{T}^{(c)}/\nu_{T}^{(0)}=20$
in the model. The same parametrization is used for the eddy diffusivity.
Figure~\ref{fig1} shows the global hydrodynamic models M1 and M2
(Table~1) with one and two circulation cells along the radius in
the solar convection zone. Our models show good agreement with the
results of helioseismology in terms of the differential rotation profile. 
The maximum difference is less than 10
nHz. Although, the magnitude of the radial shear in tachocline is
about twice higher than in the observations, and in the subsurface
shear layer, it is about twice smaller.

\subsection{Dynamo model}

Evolution of the large-scale axisymmetric magnetic field, $\overline{\mathbf{B}}$,
is governed the mean-field induction equation \citep{Krause1980},
\begin{equation}
\mathrm{\partial_{t}\overline{\mathbf{B}}=\boldsymbol{\nabla}\times\left(\boldsymbol{\mathcal{E}}+\mathbf{\overline{U}}\times\overline{\mathbf{B}}\right)},\label{eq:mfe-1}
\end{equation}
where $\boldsymbol{\mathcal{E}}=\left\langle \mathbf{u\times b}\right\rangle $
is the mean electromotive force with $\mathbf{u}$ and $\mathbf{b}$
standing for fluctuating turbulent velocity and magnetic field respectively.
We employ the mean electromotive force in the form: 
\begin{equation}
\mathcal{E}_{i}=\left(\alpha_{ij}+\gamma_{ij}\right)\overline{B}_{j}-\eta_{ijk}\nabla_{j}\overline{B}_{k}.\label{eq:EMF-1-1}
\end{equation}
where symmetric tensor $\alpha_{ij}$ models generation of the large-scale
magnetic field by the $\alpha$-effect; antisymmetric tensor $\gamma_{ij}$
controls the mean drift of the large-scale magnetic fields in turbulent
medium; the tensor $\eta_{ijk}$ governs the anisotropic turbulent
diffusion. The reader can find further details about the dynamo model
in \citet{Pipin2018b}.

The anisotropic diffusion plays a particular role in our model because
it affects overlap of the magnetic cycles, and the extended mode of
the dynamo waves \citep{Pipin2014}. We employ the anisotropic diffusion
tensor following the formulation of \citet{Pipin2008a} and \citet{Pipin2014}:
\begin{eqnarray}
\eta_{ijk} & = & 3\eta_{T}\left\{ \left(2f_{1}^{(a)}-f_{2}^{(d)}\right)\varepsilon_{ijk}+2f_{1}^{(a)}\frac{\Omega_{i}\Omega_{n}}{\Omega^{2}}\varepsilon_{jnk}\right\} \label{eq:diff}\\
 & + & \eta_{A}\phi_{1}\left(g_{n}g_{j}\varepsilon_{ink}-\varepsilon_{ijk}\right)\nonumber 
\end{eqnarray}
where $\mathbf{g}$ is the radial unit vector, $\eta_{T}$ is the
magnetic diffusion coefficient, $\eta_{A}=a\eta_{T}$, and $a$ is
a parameter of the turbulence anisotropy. The quenching functions
$f_{1,2}^{(a,d)}$ and $\phi_{1}$ are formulated by \citet{Pipin2014}.

The $\alpha$-effect takes into account kinetic and magnetic helicities
in the following form: 
\begin{eqnarray}
\mathrm{\alpha_{ij}} & \mathrm{=} & \mathrm{C_{\alpha}\frac{\nu_{T}}{Pm_{T}}\psi_{\alpha}(\beta)\alpha_{ij}^{(H)}+\alpha_{ij}^{(M)}\frac{\overline{\chi}\tau_{c}}{4\pi\overline{\rho}\ell^{2}}}\label{alp2d-2}
\end{eqnarray}
where $\mathrm{C_{\alpha}}$ is a free parameter which controls the
strength of the $\alpha$-effect due to turbulent kinetic helicity;
tensors $\mathrm{\alpha_{ij}^{(H)}}$ and $\mathrm{\alpha_{ij}^{(M)}}$
express the kinetic and magnetic helicity parts of the $\alpha$-effect,
respectively; $\mathrm{Pm_{T}}=\nu_{T}/\mathrm{\eta_{T}}$ is the
turbulent magnetic Prandtl number, and $\overline{\chi}=\left\langle \mathbf{a}\cdot\mathbf{b}\right\rangle $
($\mathbf{a}$ and $\mathbf{b}$ are fluctuating parts of the magnetic
field vector-potential and magnetic field vector). Both, $\mathrm{\alpha_{ij}^{(H)}}$
and $\mathrm{\alpha_{ij}^{(M)}}$, depend on the Coriolis number.
Function $\psi_{\alpha}(\beta)$ controls the so-called ``algebraic''
quenching of the $\alpha$-effect, where $\mathrm{\beta=\left|\overline{\mathbf{B}}\right|/\sqrt{4\pi\overline{\rho}u_{c}^{2}}}$,
$\mathrm{u_{c}}$ is the RMS of the convective velocity. It is found
that $\psi_{\alpha}(\beta)\sim\beta^{-3}$ for $\beta\gg1$. The $\alpha$-effect
tensors, $\mathrm{\alpha_{ij}^{(H)}}$ and $\mathrm{\alpha_{ij}^{(M)}}$,
are given by \citet{Pipin2018b}. Evolution of the small-scale magnetic
helicity, $\overline{\chi}=\left\langle \mathbf{a}\cdot\mathbf{b}\right\rangle $,
is governed by the magnetic helicity conservation law \citep{Pipin2011}.

We assume that the large-scale magnetic field is vanished: $B=0,\,A=0$;
and that the normal component of the magnetic field and the tangential
components of the mean electromotive force are continuous at the interface
between the tachocline and the convection zone. Following ideas of
\citet{Moss1992} and \citet{Pipin2011a}, we formulate the top boundary
condition in the form that allows penetration of the toroidal magnetic
field to the surface: 
\begin{eqnarray}
\delta\frac{\eta_{T}}{r_{e}}B\left(1+\left(\frac{\left|B\right|}{B_{\mathrm{esq}}}\right)^{2}\right)+\left(1-\delta\right)\mathcal{E}_{\theta} & = & 0,\label{eq:tor-vac}
\end{eqnarray}
where $r_{e}=0.99R_{\odot}$, and parameter $\delta=0.99$ and $B_{\mathrm{esq}}=50$G.
The magnetic field potential outside the domain is 
\begin{equation}
A^{(vac)}\left(r,\mu\right)=\sum a_{n}\left(\frac{r_{e}}{r}\right)^{n}\sqrt{1-\mu^{2}}P_{n}^{1}\left(\mu\right).\label{eq:vac-dec}
\end{equation}

The influence of magnetic field and meridional circulation on the
angular velocity profiles can be characterized by the local forces
caused by magnetic feedback on the turbulent stresses, $\hat{\mathbf{T}}$.
We introduce the following notations for components of the force per
unit mass: 
\begin{eqnarray}
F_{I} & = & -\frac{1}{\overline{\rho}r\sin\theta}\boldsymbol{\nabla\cdot}\left(r\sin\theta\overline{\rho}\hat{\mathbf{T}}_{\phi}\left(\boldsymbol{B}=0\right)\right)\label{fi}\\
F_{\ell} & = & -\frac{1}{\overline{\rho}r\sin\theta}\boldsymbol{\nabla\cdot}\left(r\sin\theta\overline{\rho}\left\{ \hat{\mathbf{T}}_{\phi}-\hat{\mathbf{T}}_{\phi}\left(\boldsymbol{B}=0\right)\right\} \right),\label{ftb}\\
F_{L}^{(t)} & = & \frac{1}{\overline{\rho}r\sin\theta}\boldsymbol{\nabla\cdot}\left(r\sin\theta\frac{\overline{\mathbf{B}}\overline{B}_{\phi}}{4\pi}\right),\label{fb}\\
F_{L}^{(p)} & = & \frac{1}{4\pi\overline{\rho}}\left(\overline{\mathbf{B}}\boldsymbol{\cdot\nabla}\right)\left(\boldsymbol{\nabla}\times\overline{\mathbf{B}}\right)_{\phi}-\frac{1}{4\pi\overline{\rho}}\left(\left(\boldsymbol{\nabla}\times\overline{\mathbf{B}}\right)\boldsymbol{\cdot\nabla}\right)\overline{\mathbf{B}}{}_{\phi}\label{flp}\\
 & + & \mathrm{\frac{1}{\overline{\rho}^{2}}\left[\!\!\boldsymbol{\nabla}\overline{\rho}\times\left(\!\!\boldsymbol{\nabla}\frac{\overline{\mathbf{B}}^{2}}{8\pi}-\frac{\left(\overline{\mathbf{B}}\boldsymbol{\cdot\nabla}\right)\overline{\mathbf{B}}}{4\pi}\!\right)\!\!\right]_{\phi}}\nonumber \\
F_{U} & = & -\frac{1}{\overline{\rho}r\sin\theta}\boldsymbol{\nabla\cdot}\left(r^{2}\sin^{2}\theta\overline{\rho}\Omega\mathbf{\overline{U}^{m}}\right),\label{fU}\\
F_{H} & = & -\frac{1}{\overline{\rho}r\sin\theta}\boldsymbol{\nabla\cdot}\left(r\sin\theta\overline{\rho}\hat{\mathbf{T}}_{\phi}^{\left(\Lambda\right)}{\left(H^{(0)}\right)}\right),\label{fH}
\end{eqnarray}
Here, $F_{I}$ represents the hydrodynamic inertial force of the solar
differential rotation. Contribution $F_{\ell}$ is caused by the Maxwell
stresses of the small-scale magnetic fields. This effect was computed
analytically in the previous studies cited above. It is zero in absence
of magnetic field. $F_{L}^{(t,p)}$ stands for the of the large-scale
magnetic field, and $F_{U}$ is the azimuthal force due to the effect
of the meridional circulation. Contribution $F_{H}$ describes the
dynamo-induced latitudinal angular momentum flux by the $\Lambda$-effect.
In the dynamo theory, see, e.g. \citet{Kueker1996}, the magnetically
induced $\Lambda$ effect is usually denoted by $H^{(0)}$ (see details
in Appendix). The $H^{(0)}$ term results from the small-scale Maxwell
stresses $\overline{b_{\theta}b_{\phi}}/4\pi\overline{\rho}$, which
stem from perturbations of the large-scale magnetic field by turbulence.
This effect was calculated analytically in the above cited papers.
It was found that $H^{(0)}$ can be represented as the sum of two
contributions, 
\begin{equation}
H^{(0)}=H^{(0,a)}+H^{(0,\rho)},\label{h0}
\end{equation}
where $H^{(0,a)}$ originates from anisotropy of convective turbulence
\citep{Kichatinov1988}, and $H^{(0,\rho)}$ results from the density
stratification \citep{Kueker1996,Pipin1999}. The signs of $H^{(0,a)}$
and $H^{(0,\rho)}$ are opposite. Unlike the other contributions of
the $\Lambda$-effect, $H^{(0)}$ grows with the increase of energy
of the large-scale magnetic field when the field is weak, $\beta<1$.
Further theoretical details are given in Appendix A. Another promising
source of the dynamo induced $\Lambda$ effect stems from the convective
heat flux \citep{Rogachevskii2018}. We postpone discussion of this
effect for future papers. 

The dynamo cycle results in variations of large-scale torque forces.
Below we discuss properties of the stationary phase of the numerical
solution. We consider deviations of the force components from their
mean values (time averages). In addition, there is an effect of variations
of the entropy gradient because the magnetic field affects the heat
transport by means of magnetic quenching of the turbulent eddy heat
conductivity, as well as the energy sinks and gains associated with
magnetic field generation and dissipation (see Eq.~\ref{eq:heat}).
Consequently, the entropy variations result in perturbations of the
Taylor-Proudman balance \citep{Durney1999} and variations of the
meridional circulation. The latter affects the magnitude and distribution
of the differential rotation in the solar convection zone, and play
important role in the mechanism of torsional oscillations.

\section{Results}

Results of our numerical experiments show that there are two necessary
conditions for appearance of the \emph{extended }22-year pattern of
the torsional oscillations. The first one is existence of the 22-year
mode in the dynamo wave pattern and the subsequent overlap of magnetic
cycles on the solar surface. The second one is magnetic quenching
of the eddy thermal conductivity. This effect results in the cyclic
modulation of the thermal wind associated with the mean entropy gradient,
and, correspondingly, the Taylor-Proudman balance and the meridional
circulation. The latter affects the torsional oscillations by means
the forces $F_{U}$ and $F_{I}$. Without the above two conditions
the model can only reproduce the 11-year torsional oscillation pattern
that is known from other mean-field and numerical models that can
be found in the literature (e.g., \citealt{Kueker1996,Pipin1999,Kuker1999,Pipin2004,Covas2000,Rempel2006,Guerrero2016a,Kapyla2016}).

Table \ref{tab:Models} shows a set of our model runs. We use the
same dynamo parameters as in our previous study \citep{Pipin2018b}.
They are the magnetic Prandtl number, $\mathrm{Pm_{T}}={\displaystyle \frac{\nu_{T}}{\eta_{T}}}=10$,
the $\alpha$-effect coefficient, $C_{\alpha}=0.04$, and the magnetic
Reynolds number, $R_{m}=10^{6}$. The coefficient, $C_{\alpha}$,
is chosen about 5 percents above the dynamo threshold. The magnitude
of the $\alpha$-effect is about 1 m/s, and it changes sign near the
bottom of the convection zone from positive to negative in the northern
hemisphere, and from negative to positive in the southern hemisphere.
The dynamo model includes a simplified implementation of the tachocline
region below the convection zone. The $\alpha$-effect in this region
is suppressed, and it is not involved in turbulent generation of large-scale
magnetic field (cf, e.g., \citealp{Ruediger1995}). The tachocline
plays a role of storage for the large-scale toroidal magnetic field,
which penetrates from the convection zone. It increases efficiency
of the distributed dynamo operating in the solar convection zone \citep{Guerrero2016}.

Models M1 and M2 are fully nonlinear models for the case of one and
two meridional circulations cells along the radius in the northern
and southern segments of the convection zone. These models show the
22-yr dynamo modes in evolution of the near surface toroidal magnetic
fields. In model M3, we switch on the anisotropic eddy-diffusivity
parameter. It reduces overlap of the subsequent cycle. In this model,
we consider the one-cell meridional circulation case, like in most
runs presented in this paper, because the results for the single-
and double-cell meridional circulations are qualitatively similar.
Also, it was found that, in general, the models which involve the
$F_{H}$-force component $H^{(0,\rho)}$ that stems from the density
stratification give better agreement with observations than the models
with the force component, $H^{(0,a)}$, resulting from the anisotropy
of the background convection. Therefore, we mostly show results for
the models with the $H^{(0,\rho)}$ effect. Models M4, M5 and M6 are
calculated to study effects of the large- and small-scale Lorentz
force, as well as the difference between contributions of $H^{(0,\rho)}$
and $H^{(0,a)}$ in the dynamo induced $\Lambda$ effect, $H^{(0)}$.
In model M7, we neglect the magnetic quenching of the eddy thermal
conductivity. This substantially reduces the mean entropy variations
in the dynamo cycle, and correspondingly reduces variations of the
meridional circulation and their effect on the angular momentum balance.
Table \ref{tab:Models} summarizes descriptions of the runs. 

Table \ref{tab:Models-r} presents some model characteristics that
are relevant for our discussion of the origin of the extended mode
of the torsional oscillations. The dynamo cycle period is determined
from the polar magnetic field reversals. To characterize the dynamo
cycle overlap we calculate the relative areas occupied by the toroidal
magnetic flux of one sign at the bottom of the subsurface shear layer
(SSL), $\mathrm{S^{(T)},}$ in the subsurface shear layer ($r=0.9-0.99R$),
$\mathrm{S_{SSL}^{(T)}}$, and inside the bulk of the convection zone,
$\mathrm{S_{CZ}^{(T)}}$: 
\begin{eqnarray}
\mathrm{S^{(T)}} & = & \int_{0}^{1}\mathrm{sign}\left(\overline{\mathbf{B}}{}_{\phi}\left(\mathrm{r=r_{s}}\right)\right)\mathrm{d\mu},\label{eq:st}\\
\mathrm{S_{SSL}^{(T)}} & = & \frac{1}{\left(R^{2}-r_{s}^{2}\right)}\int_{0}^{1}\mathrm{\int_{r_{s}}^{R}sign}\left(\overline{\mathbf{B}}{}_{\phi}\right)\mathrm{r\mathrm{dr}\mathrm{d}\mu},\label{eq:stssl}\\
\mathrm{S_{CZ}^{(T)}} & = & \frac{1}{\left(R^{2}-r_{b}^{2}\right)}\int_{0}^{1}\int_{r_{b}}^{R}\mathrm{sign}\left(\overline{\mathbf{B}}{}_{\phi}\right)r\mathrm{dr}\mathrm{d}\mu,\label{eq:stcz}
\end{eqnarray}
where $r_{s}=0.9\,R$, and $r_{b}=0.728\,R$. These parameters reach
maximum value when the whole hemisphere is occupied by the magnetic field
of one sign. In particular, when some of these parameters are equal
to $\pm1$, the whole hemisphere is occupied by the magnetic field
of one sign. The absolute value of the parameter less than 1 means
the magnetic cycles overlap. We define the time delay between the
subsequent dynamo cycles by the life-time of the state, $S^{(T)},S_{SSL}^{(T)},S_{CZ}^{(T)}=\pm1$.
Similarly to \citet{Sten1992}, we measure the duration of the extended
dynamo mode from the butterfly diagram of the radial magnetic field
on the surface. The starting position of the extended mode corresponds
to beginning of the polar and equatorial branches at approximately 55$^{\circ}$
latitude. In addition, Table \ref{tab:Models-r} gives magnitudes
of the torsional oscillations and zonal acceleration on the surface, the
duration of the equatorward part of the extended mode of the torsional
oscillations, and the duration of the polar branch of the torsional
oscillations.

\subsection{Full models}

We start with describing the full dynamo models that take into account
all nonlinear effects, and reproduce the extended mode of the torsional
oscillation. Figure \ref{m0-tl} shows results for the magnetic field
evolution in model M1 at the top boundary of the convection zone together
with the corresponding evolution of dynamo-induced zonal variations
of the rotational velocity and zonal acceleration, as well as contributions
of the zonal force components. The large-scale toroidal magnetic field
at the bottom of the subsurface shear layer varies with the magnitude
of 1.5~kG. In the subsurface shear layer, the dynamo wave of the
toroidal magnetic field starts at about $60^{\circ}$ latitude, approximately
1-2 years after the end of the previous activity cycle. The magnitude
of the toroidal magnetic field at this latitude is about 4~G. The
wave propagates toward the equator in $\sim22$ years. The polar and
equatorial branches of the dynamo waves are almost completely overlapped
in time. The radial magnetic field reveals the extended mode as well.
Its evolution is in agreement with the observational results of \citet{Stenflo1988}
and \citet{Sten1992}.

The dynamo wave forces variations of the angular velocity and meridional
circulation. It is seen in Fig.~\ref{m0-tl}e that the induced zonal
acceleration is $\sim2-4\times10^{-8}~$m$\,$s$^{-2}$, which is
in agreement with the observational results of \citet{Kosovichev2019}.
However, the individual force contributions are by more than an order
of magnitude stronger than their combined action. Another interesting
finding is that two components of the azimuthal force show the extended
22-year modes. They are: $F_{U}$, associated with variations of the
meridional circulation, and the inertial force, $F_{I}$. In model
M1, the large-scale Lorentz force, $F_{L}^{(t)}$ displays an extended
pattern corresponding to 16 years of drifting to the equator. The
equatorial branches of forces, $F_{U}$ and $F_{I}$, are nearly in
balance. The effect of magnetic field on the turbulent stress does
not show the extended cycle, or it is strongly quenched at high latitudes.
However, the total force, $F_{\mathrm{TOT}}=F_{\ell}+F_{I}+F_{L}^{(t)}+F_{U}$,
clearly shows the extended mode (Fig.~\ref{m0-tl}h). Its time-latitude
pattern corresponds to the zonal acceleration evolution. It is seen
that different parts of the zonal acceleration diagram correspond
to the different excitation forces. For example, the polar brunch
is in phase with $F_{U}$ and $F_{L}^{(t)}$, at mid latitudes the
zonal acceleration is in phase with $F_{L}^{(t)}$, at low latitudes
it is in phase with $F_{\ell}$, and the equatorial part is in phase
with $F_{U}$. It is important that the model takes into account force
$F_{U}$, resulting from the nonlinear effects of forces $F_{L}^{(t,p)}$
and $F_{\ell}$ driving the torsional oscillations. Force $F_{I}$
is the restoring inertial force. The results for model M2 with two
meridional circulations cells are qualitatively similar to M1. 

Figure~\ref{trm3} shows the time-radius diagrams of the
evolution of the large-scale magnetic field and flows in model M1 
for four latitudes, 10$^{\circ}$, 30$^{\circ}$, 45$^{\circ}$, and 60$^{\circ}$.
The large-scale magnetic field migrates in two directions: upward
in the main part of the convection zone, and downward near the bottom
of the convection zone and in the tachocline. Generally, direction
of propagation of the torsional oscillations corresponds to propagation
of the dynamo wave at both, the high and low latitudes. At 60$^{\circ}$
latitude, the inclination of the torsional wave pattern in the time-radius
diagram is small, and variations of the zonal acceleration are synchronized
by the phase at all depths of the solar convection zone. At lower
latitudes, the torsional waves propagate upward from the bottom of
the convection zone, and form almost stationary oscillatory patterns
in the subsurface layer, the depth of which increases with the latitude
decrease. These effects qualitatively correspond to the results of
our recent helioseismology analysis \citep{Kosovichev2019}.

Figure \ref{czM1} shows the evolution of the large-scale magnetic
field, flows, the dynamo-induced forces variations of convective flux
$\delta F_{c}/F_{\odot}$ (where $F_{\odot}$ is the total energy
flux), and variations the Taylor-Proudman balance (TPB) inside the
convection zone in model M1. The TPB includes all terms of the right
hand side of Eq.~(\ref{eq:vort}) except the advection term. The
results for model M2 are very similar. Both models show the qualitatively
similar results despite the different meridional circulation structure
in the convection zone. The secondary deep clockwise meridional cell
of magnitude 1~m/s is weak, and does not significantly affects the
dynamo wave propagation in model M2. In both models, M1 and M2, the
maximum of the toroidal magnetic field strength near the bottom of
the convection zone is about 5~kG. The strength of the toroidal magnetic
field drops to about 1~kG at $0.9R$. At low latitudes, the dynamo
wave propagates to the equator, upward in the upper part of the convection
zone and downward in the tachocline. At high latitudes, there is a
poleward propagating branch, which has a rather weak signature at
the surface. It is also seen in the time-latitude diagrams of the
toroidal magnetic field in the subsurface layers, (see, Fig.~\ref{m0-tl}).

A new dynamo cycle starts near the bottom of the convection zone {
at about 60 degrees latitude} in the region subjected to the zonal
{ deceleration}. The acceleration pattern in the upper part of the
convection zone drifts toward equator following the propagation of
the dynamo wave (see, Fig.~\ref{czM1}a,b). Figures~\ref{czM1}d,e
shows that the zonal deceleration is ahead of the toroidal magnetic
field due to the combined effect of the large-scale Lorentz force,
$F_{L}^{(t)}$, and the dynamo induced $\Lambda$-effect, $F_{H}$.
Notably, the $F_{H}$ force migrates closer to the equator than the
$F_{L}^{(t)}$ force. The forces $F_{U}$ (Fig.~\ref{czM1}b) and
$F_{I}$ (Fig.~\ref{czM1}e) are opposite in sign, and they have
similar latitudinal structures. We see that in the latitude range
from 50$^{\circ}$ to 60$^{\circ}$, where the extended dynamo mode
is initiated, the acceleration is provided by the inertia force, $F_{I}$
(Figs.~\ref{czM1}b and e). Also, we see that, during a half of the
extended torsional oscillation cycle, effects of $F_{U}$, $F_{L}$
and $F_{H}$ are synchronized in the subsurface layer of the convection
zone. The polar branch of the torsional oscillations in model M1 is
due to effects of the Lorentz force and variations of the meridional
circulation. The models show weak meridional circulation variations
in the main part of the convection zone, where its magnitude is about
10--20 cm/s. The surface variations of the meridional circulation
in the dynamo cycle are about 1 m/s. Figure~\ref{czM1}c shows that
the direction of the excited meridional flow, which migrates to the
equator ahead of the toroidal magnetic field, is counter-clockwise
(poleward on the surface), and that it is clockwise for the flow behind.
This corresponds to the meridional flow variations converging towards
the activity belts in agreement with results of local helioseismology
\citep{Komm2012,Zhao2014,Kosovichev2016}. 

Figure~\ref{czM1}f shows variations of the convective flux $\delta F_{c}/F_{\odot}$,
where $F_{\odot}$ is the total energy flux, and the relative variations
the Taylor-Proudman balance (TPB) inside the convection zone in model
M1. We see that the large-scale toroidal magnetic field results in
a reduction of the convective flux. This phenomenon, called the magnetic
shadow effect, was discussed earlier by, e.g., \citet{Brandenburg1992}
and \citet{Pipin2004}, and is usually considered in the problem of
the solar-cycle luminosity variation. In the model, the magnetic shadow
effect induces variation of the latitudinal gradient of the mean entropy.
This results in perturbation of the Taylor-Proudman balance and variations
of the meridional circulation. In agreement with other studies, e.g.
of \citet{Durney1999}, \citet{Rempel2006} and \citet{Miesch2011a},
the variations of TPB are concentrated near the boundaries of the
convection zone, and correlate well with $\delta F_{c}/F_{\odot}$.
Therefore, the magnetic shadow effect is the main source of the TPB
and meridional circulation perturbations in our model. This was also
confirmed in separate runs when we switched off all other source of
the TPB variations in Eq.~(\ref{eq:vort}). 

Interestingly, a better agreement with the observed extended pattern
of the torsional oscillations is found in the special cases of model
M1, which are represented by models M4 and M5 in Table~\ref{tab:Models}.
In model M4, the influence of the magnetic field on the angular momentum
is restricted to effect of the large-scale Lorentz force, $F_{L}^{(t)}$,
and $F_{\ell}$, associated with the small-scale Maxwell stresses
$\overline{b_{i}b_{j}}/4\pi\overline{\rho}$ is neglected. In model
M5, which is probably in the best agreement, the $F_{L}^{(t)}$ effect
is neglected, and the torsional oscillations are driven by $F_{\ell}$.
Model M6 uses the same approach as M5 but it employs the dynamo-induced
$\Lambda$-effect of a different origin (see, Table \ref{tab:Models}).
This model does not show the extended mode of the torsional oscillation. 

\subsection{Effect of the extended dynamo mode and cycles overlap}

Both models, M1 and M2, (with the single- and double-cell meridional
circulation) show the extended mode of the torsional oscillations.
It seems natural to relate this effect with extended dynamo mode which
shows up in these models. On the Sun, this dynamo mode is deduced
from the overlap of magnetic cycles \citep{Sten1992}. To quantify
the overlap in our models, we use the parameters of the toroidal magnetic
field distributions (Eqs.~\ref{eq:st}-\ref{eq:stcz}). To demonstrate
the role of the extended dynamo mode and the cycle overlap, we consider
model M3 with anisotropy of the turbulent eddy-diffusivity, following
results of \citet{Pipin2014}. The anisotropy is controlled by parameter
$\eta_{A}$, see Eq.~(\ref{eq:diff}). In model M3, we put $\eta_{A}=2\eta_{T}$.
This decreases the overlap of the subsequent cycles. Figure \ref{fig:st}
shows evolution of the parameters, $\mathrm{S^{(T)}}$, $\mathrm{S_{SSL}^{(T)}}$,
and $\mathrm{S_{CZ}^{(T)}}$, for models M1, M2 and M3. Figure \ref{fig:st}(a)
shows that, at the bottom of the subsurface shear layer, the dynamo
waves of the toroidal magnetic field are not fully overlap in all
models, and that the time delay between the end of a magnetic cycle
at the equator, and the start of a new cycle at high latitudes is
about 1.2 years in model M1, 1.1 years in model M2, and 2.5 years
in model M3. However, in terms of the magnetic flux integrated over
the subsurface shear layer or over the bulk of the convection zone,
we find that in models M1 and M2 the dynamo waves are overlapped because
$\mathrm{S_{SSL}^{(T)}}$, and $\mathrm{S_{CZ}^{(T)}}\sim0.96$ .
Model M3 shows a time delay of about 1.5 year for the dynamo waves
in the convection zone (Fig.~\ref{fig:m13}a). 

 The delay between the subsequent cycles in the subsurface shear layer
results in an interruption of the torsional oscillation pattern in
the equatorial regions (Fig \ref{fig:m13}b). The sum of the forces,
$F_{I}+F_{U}$, that form the main balance and drive the extended
mode of the torsional oscillations in models M1 (Fig.~\ref{fig:m13}d)
and M2, but not in model M3 (Fig~\ref{fig:m13}c). Therefore, the
decrease of the dynamo-cycle overlap breaks the phase continuity in
the balance of forces driving the torsional oscillations. The cycle
overlap can interpreted as a measure of penetrations of the toroidal
magnetic field from the deep convection zone into the subsurface shear
layer in course of the magnetic cycle. The additional anisotropy of
magnetic diffusivity in model M3 disperses magnetic field in the radial
direction \citep{Pipin2014}, and, thus, reduces penetration of the
toroidal from the convection zone to the surface.\textbf{ }

\subsection{Magnetic perturbations of the convective heat flux and meridional
circulation}

Our results presented in the previous sections, showed that the dynamo-induced
variations of the meridional circulation play a significant role in
the total force balance driving the torsional oscillations. 
This effect results from the torque induced by the large-scale flows,
$-\boldsymbol{\nabla\cdot}\left(r\sin\theta\overline{\rho}\overline{U}_{\phi}\mathbf{\overline{U}^{m}}\right)$,
where $\overline{U}_{\phi}=r\sin\theta\Omega$, and $\mathbf{\overline{U}^{m}}$
is the meridional circulation velocity. Generally, $\left|\overline{U}_{\phi}\right|\gg\left|\mathbf{\overline{U}^{m}}\right|$.
The dynamo-induced perturbations of the angular velocity are of the
same order of magnitude as the perturbation of the meridional circulation.
Therefore, the leading term of the torque driving the torsional oscillations
can be written as: $-\boldsymbol{\nabla\cdot}\left(r\sin\theta\overline{\rho}\overline{U}_{\phi}\mathbf{\delta\overline{U}^{m}}\right)$,
where $\mathbf{\delta\overline{U}^{m}}$ is the dynamo-induced perturbation
of the meridional circulation velocity. In the middle of the convection
zone: $\mathbf{\delta\overline{U}^{m}}\sim$10~cm/s, $\overline{U}_{\phi}\sim2.5$~km/s.
Therefore, the effect of this perturbation is comparable with the
speed of the torsional oscillations, 5-10 m/s. 

Variations of meridional circulation are caused by perturbations of
the Taylor-Proudman balance (TPB). Results for model M1 show that,
near the surface, perturbations of the TPB correlate with variations
of the convective flux, $\delta F_{c}/F_{\odot}$, which are caused
by quenching of the convective heat flux by the large-scale magnetic
field. The impact of this effect on the torsional oscillations is
demonstrated by comparison of models M1 and M7 (Fig.~\ref{fig:th}).
In model M7, the magnetic quenching of convective flux is neglected.
This model does not show the extended mode of torsional oscillations.
Variations of the TPB and meridional circulation in model M7 are substantially
smaller in magnitude than in model M1 and they disagree with observations.
The time-latitude diagram of the meridional circulation variations
in model M1 is in a good agreement with observations \citep{Komm2012,Zhao2014,Kosovichev2016}.
To check the role of variations of the meridional circulation we made
a special run, in which we suppressed the variations, and obtained
the results similar to model M7 (Fig.~\ref{fig:th}a), without the
extended torsional oscillation pattern.

In model M8, we switch off effects of the large-scale Lorentz force,
$F_{L}^{(t,p)}=0$, and the Maxwell stresses contribution of the small-scale
magnetic fields, $F_{\ell}=0$, and $F_{H}=0$. In this case, the
torsional oscillation are driven by magnetic quenching of the convective
heat flux. This effect was previously discussed using the heuristic
arguments by \citet{Spruit2003} and \citet{Rempel2006}. In our models,
we employ analytical results of the mean-field theory using expressions
for the thermal eddy conductivity tensor, given by Eq.~(\ref{eq:htfl})
for the case of rotating and magnetized fluid \citep{Kitchatinov1994,Pipin2018b}.
The time-latitude diagram of the toroidal magnetic field, the azimuthal
velocity acceleration and variations of the meridional circulation
are shown in Figure \ref{m8tl}, and the snapshots for the different
phases of the magnetic cycle are shown in Figure \ref{m8sn}. This
model shows the extended cycle of the torsional oscillations. Except
the polar region parts, the torsional oscillation pattern is similar
to results of the model M4. The polar parts, in turn, are qualitatively
similar to model M5. Like in model M1, variations of the meridional
oscillations are in agreement with the results of observations. The results
for this model show that the effect of large-scale magnetic field
on the convective heat transport is among significant drivers of the
solar torsional oscillations.

\section{Discussion and conclusions}

Our paper addressed the question of the origin of the solar torsional
oscillations, and, in particular, its extended 22-year mode. We presented
a non-linear mean-field dynamo model that self-consistently describes
the differential rotation and meridional circulation, as well as the
generation and transport of the large-scale magnetic field. For the
first time, the model explains the phenomenon of the extended solar
cycle observed in the torsional oscillation pattern, and reveals the
balance of forces that drive the torsional oscillations.

The model is based on the mean-field magnetohydrodynamics theory of
turbulent generation of large-scale magnetic field, as well as turbulent
heat and angular momentum transport. We employ the reference state
computed from the standard evolutionary model of the Sun, and use
turbulent parameters of the convective RMS velocity to calculate the
mean-field non-diffusive turbulent effects and heat transport. The
angular momentum transport in the convection zone is affected by the
large-scale Lorentz force and magnetic effects on the turbulent stress
tensor, $\hat{T}_{\mathrm{ij}}=\overline{u_{i}u_{j}}-\overline{b_{i}b_{j}}/4\pi\overline{\rho}$.
The turbulent stresses cause generation and dissipation of the large-scale
flows (see Eq.~\ref{eq:tstr}). The dynamo-generated magnetic field
results in quenching of the turbulent angular momentum fluxes as well
as the non-dissipative angular momentum flux, which in terms of the
mean-field theory is described as the $\Lambda$-effect. In the case
of the weak mean field, its amplitude is proportional to $\overline{B}^{2}$.
Our model describes the dynamo processes distributed in the convection
zone.

Previous models of the torsional oscillations were constructed by
coupling the dynamo equations with equations of the angular momentum
transport \citep[e.g.][]{Malkus1975,Kueker1996,Pipin1999,Kuker1999,Covas2000}.
However, such models did not describe the angular momentum flux induced
by variations of the meridional circulation. These variations naturally
result from perturbations of the Taylor-Proudman balance, and are
caused by the dynamo-induced variations of the centrifugal force,
variations of the ``thermal wind'' (associated with the mean entropy
gradient), magnetic quenching of the eddy-viscosity and the large-scale
Lorentz force. Dynamo-induced perturbations of the Taylor-Proudman
balance in the framework of the mean-field theory were previously
considered by \citet{Brandenburg1992} and \citet{Rempel2007ApJ}.
Therefore, it is important that the nonlinear consistent dynamo models
presented in this paper describe both, the differential rotation and
the meridional circulation, as well as their variations.

We find two essential conditions that affect excitation of the extended
mode of the torsional oscillations in the large-scale dynamo operating
in the bulk of the convection zone. They are as follows: a) the existence
of the extended mode in the dynamo wave pattern and overlap of subsequent
dynamo waves in the time-latitude diagram; b) the dynamo-cycle variations
of the meridional circulation which stem from quenching the convective
heat transport by the large-scale magnetic field. This effect is called
the magnetic shadow \citep{Stix2002}. It represents a mean-field
analog of the phenomenon observed in sunspots \citep{Kitchatinov1994}.
Condition (a) is related to the distribution of the dynamo process
inside of the convection zone. Our models show that the dynamo waves
extent over the depth of the convection zone and over latitudes from
the equator to polar regions. The magnetic forces are distributed
similarly (e.g., the snapshots of the large-scale Lorentz force, $F_{L}^{(t)}$
and the dynamo induced $\Lambda$-effect, $F_{H}$, in Fig.~\ref{czM1}c
and d.) This is different to the flux-transport models, where the
dynamo wave occupies only a particular range of depth, near and below
the bottom of the convection zone \citep[e.g.][]{Choudhuri1999,Rempel2006}.
To demonstrate the importance of this condition we calculated a dynamo
model that includes effects of anisotropy of background convective
flows on the eddy diffusivity in a parametric form \citep{Pipin2014}.
This modification restricts the latitudinal extent of the dynamo wave,
reduces the overlap of the subsequent cycles, and results in disappearance
the extended mode of the torsional oscillations. Inspection of results
for the other models (see, the Table \ref{tab:Models-r}) shows that
the cycle overlap depends on the nonlinear effects which take part
in the magnetic field evolution. 

The importance of condition (b) -- the dynamo-induced variations
of the convective flux that results in the meridional circulation
variations -- is confirmed by results of model M7 where these variations
were suppressed. Coupling with variations of the Taylor-Proudman balance,
the variations of the meridional circulation affect the amplitude
of the torsional oscillations and produce the effective drifts of
the torsional wave in radius and latitude. Our modeling results showed
that neglecting the meridional circulation variations results in disconnections
in the torsional oscillation patterns near the bottom of the convection
zone and at the surface. In our model, the greatest effect on the
TPB results from variations of the mean entropy caused by variations
of the convective heat transport. Thus, neglecting this effect, as
shown in model M7, results in no extended mode of the torsional oscillations. 

Results of our models show that the evolutionary pattern of the torsional
waves can depend on the physics involved in the effects of magnetic
field on the angular momentum transport (see, Eqs~\ref{fi}-\ref{fH}).
These effects are usually considered as the principal sources of the
11-year modulations of zonal flows in sunspot cycles. We find that
the extended mode of the torsional oscillations can be qualitatively
explained in different ways, either by combination of the magnetic
feedback on the turbulent fluxes of the angular momentum transport,
the $F_{\ell}$ and $F_{H}$ forces (called the $\Lambda$-quenching),
and the large-scale Lorentz force, $F_{L}^{(t)}$, as in model M1,
or by separate actions by the large-scale Lorentz force (model M4)
and the $\Lambda$-effect quenching (model M5). The results
of global MHD models of \cite{Guerrero2016a} and \cite{Kapyla2016}
show significance of the large-scale Lorentz force for evolution of
the high-latitude branches of the torsional oscillations. This is
also found in our models M1 and M4.

It is interesting that in both models, M4 and M5, the magnetic drivers
of the torsional oscillations are related to the forces which reduce
the large-scale latitudinal shear. Model M6 employs the dynamo induced
$\Lambda$-effect, $H^{(0,a)}$ which is caused by the anisotropy
of background convective flows. This effect works to increase the
latitudinal shear. However it does not produces the extended mode. 

\begin{figure}
\includegraphics[width=0.95\textwidth]{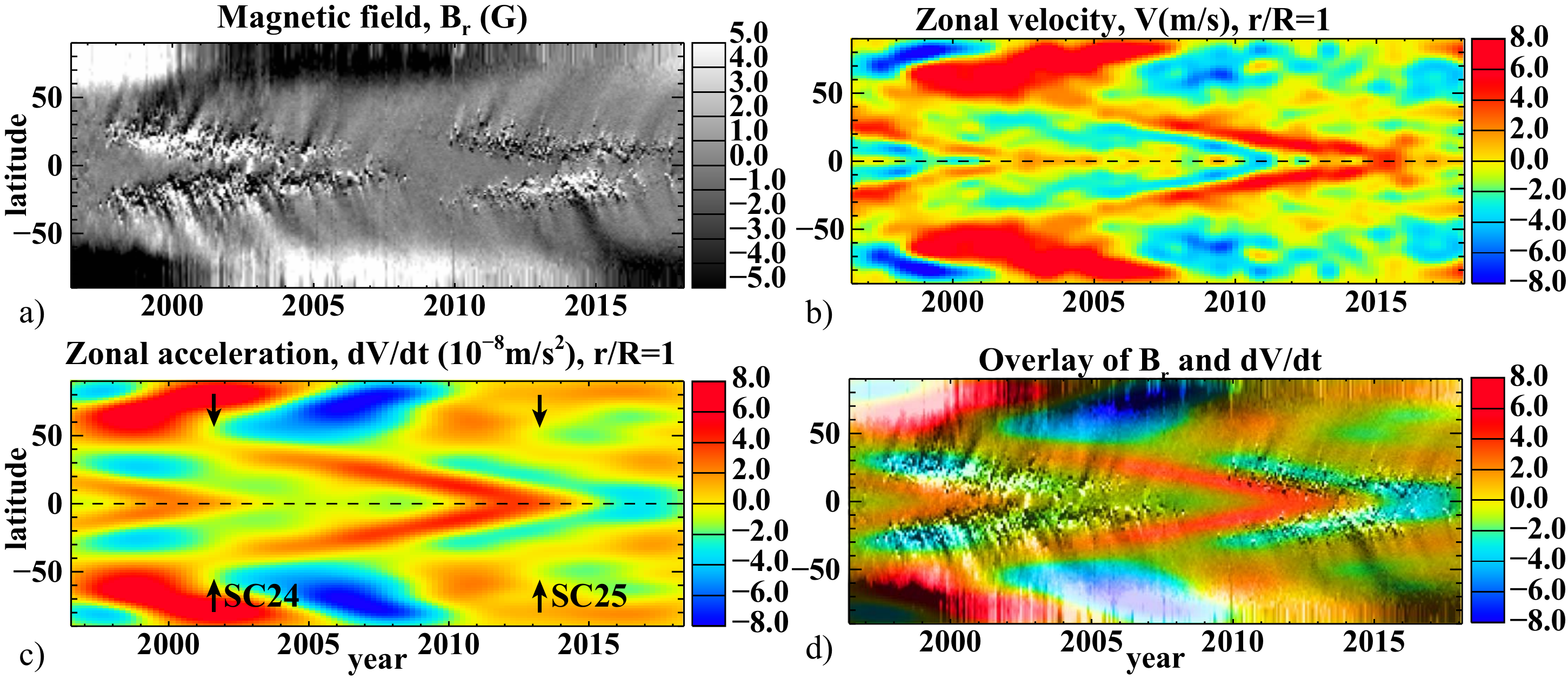} \caption{Observational results from SoHO/MDI and SDO/HMI. The time-latitude
diagrams of: a) evolution of radial magnetic field during the last
two solar cycles; b) zonal flow velocity (`torsional oscillations');
c) zonal acceleration calculated after applying a Gaussian spatial
and temporal filter, arrows indicate the start of extended solar cycles
24 and 25 at about 55 degrees latitude, defined as a starting point
of the zonal deceleration (blue ares); d) overlay of the zonal acceleration
(color image) and the radial magnetic field (gray-scale) \citep[after][]{Kosovichev2019}.}
. \label{figure01} 
\end{figure}

\begin{figure}
\includegraphics[width=0.95\textwidth]{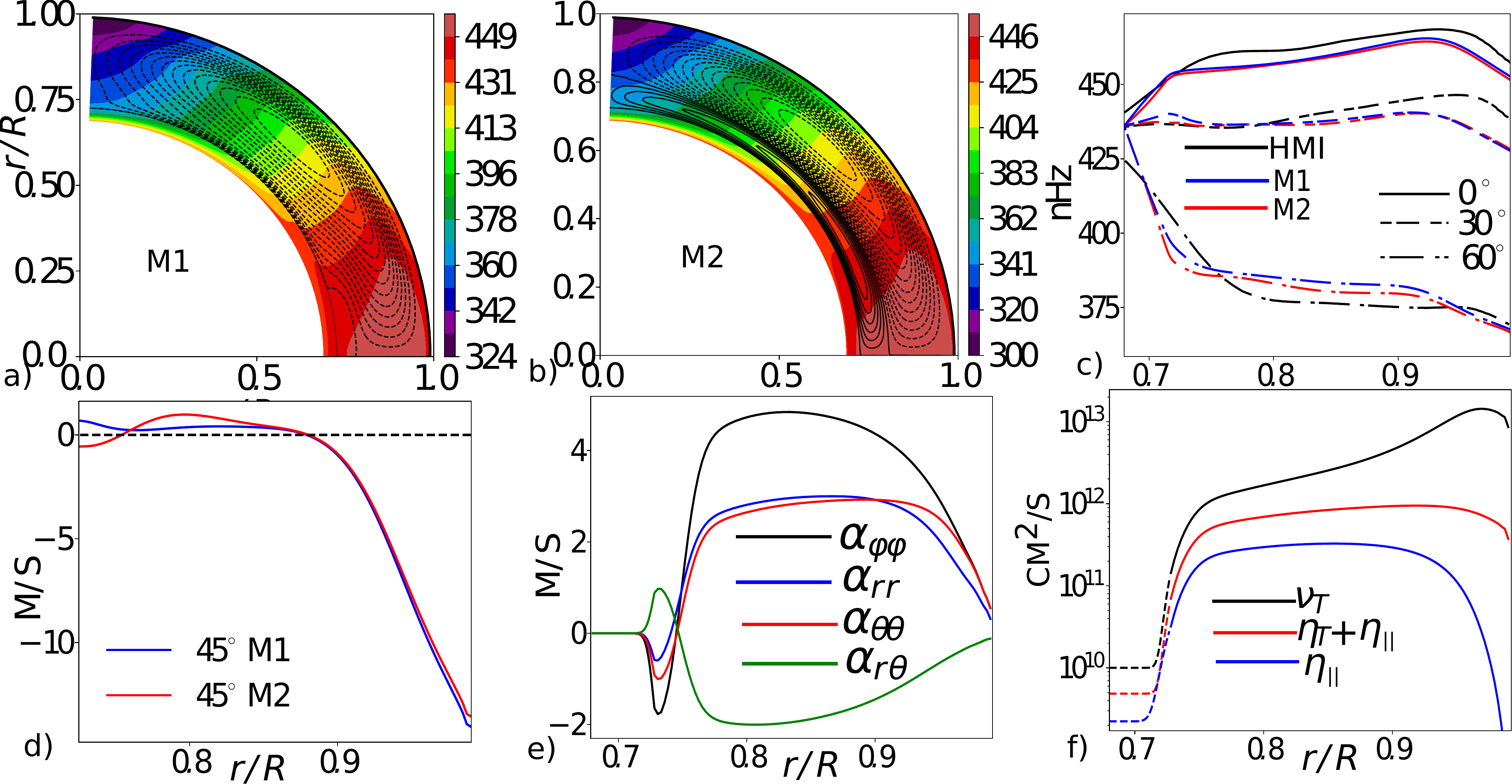} \caption{\label{fig1} The angular velocity and meridional circulation distributions
for the models specified in Table~1: a) model M1; b) model M2; c)
radial profiles of the angular velocity at 0, 30 and 60 degrees latitudes
\textbf{( blue for model M1, red - M2), and obtained from the SDO/HMI
helioseismology data archive (black);} d) radial profiles of the meridional
circulation velocity at latitude 45$^{\circ}$; e) radial profiles
of the $\alpha$-effect tensor at latitude 45$^{\circ}$; f) radial
profiles of the isotropic part of the eddy viscosity, $\nu_{T}$,
the total, $\eta_{T}+\eta_{||}$, and rotationally induced part, $\eta_{||}$,
of the eddy magnetic diffusivity.}
\end{figure}

\begin{figure}
\includegraphics[width=0.95\textwidth]{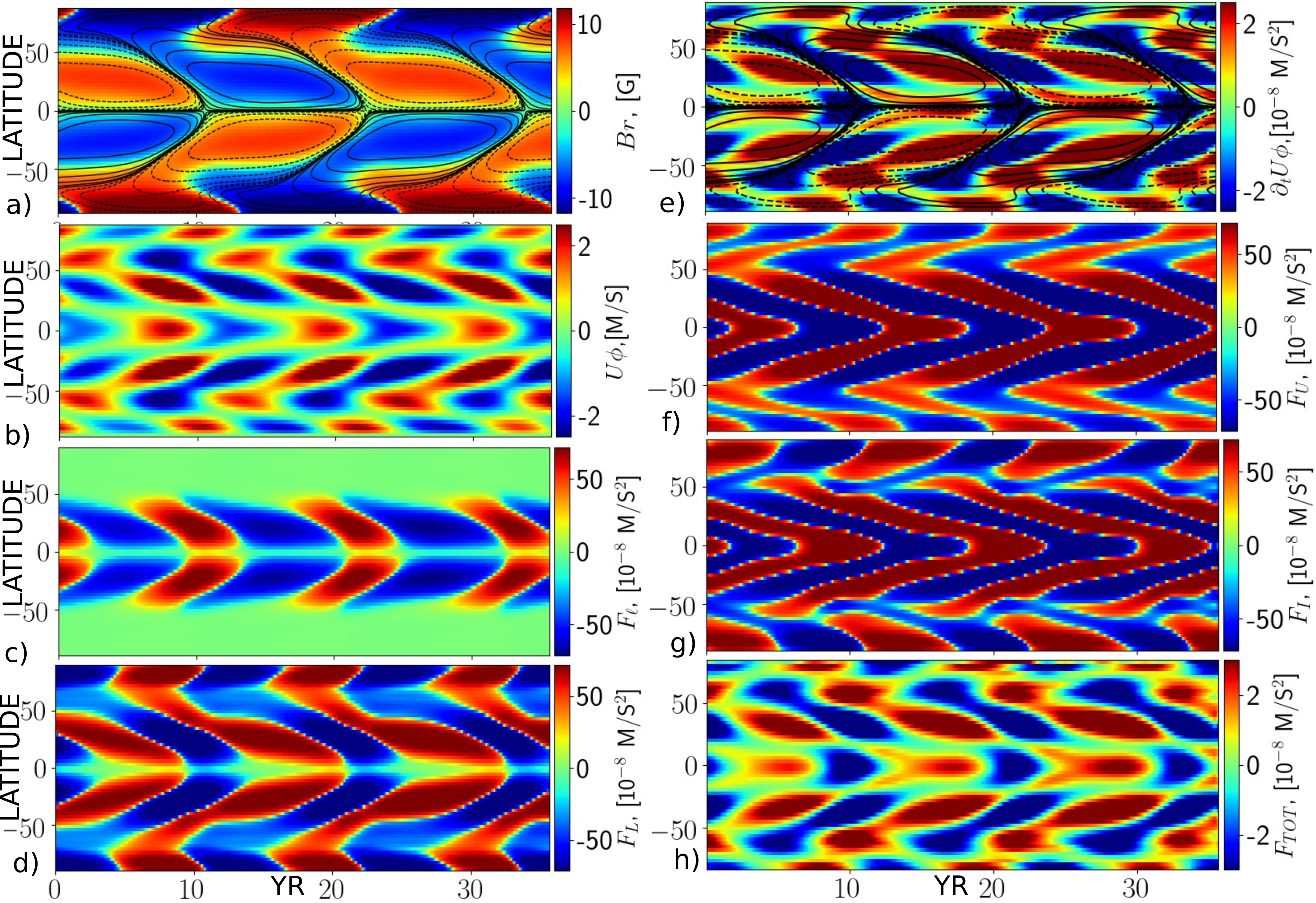} \caption{\label{m0-tl}Model M1. a) Time-latitude diagram of the radial magnetic
field at the surface (background image) and the toroidal magnetic
field at the bottom of the subsurface shear layer (SSL) $r_{s}=0.9R$,
the contour lines are plotted in range of $\pm$1kG with \textit{exponential
decrease of magnitude} to the low values of $\pm$4G; b) time-latitude
diagram of the torsional oscillations (background image) at the surface;
c) azimuthal density force from variations of the turbulent stress
at the surface, $F_{\ell}$; d) the same as c) for the large-scale
Lorentz force, $F_{L}^{(t)}$; e) time-latitude diagram of the zonal
flow acceleration (background image) at the surface and the toroidal
magnetic field in the subsurface shear layer (same as panel (a));
f) the same as in panel (d) for the ``inertial'' forces, $F_{I}$;
g) the same as in panel (d) for the effect of the meridional flow,
$F_{U}$; h) the total force at the surface, $F_{\mathrm{TOT}}=F_{\ell}+F_{I}+F_{L}^{(t)}+F_{U}$.}
\end{figure}

\begin{figure}
\includegraphics[width=0.85\textwidth]{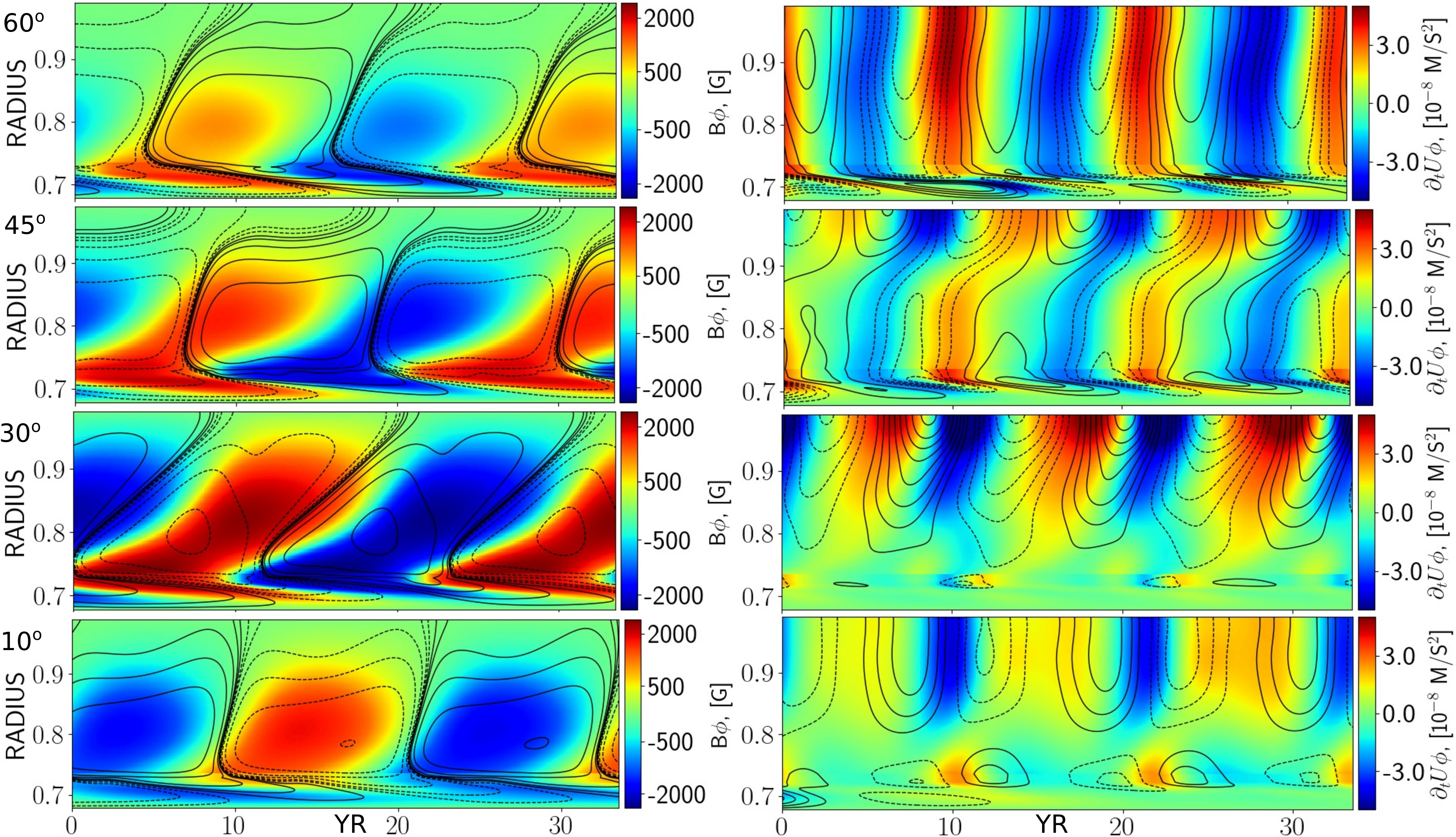}\caption{\label{trm3} Model M1. Left column shows time-radius diagrams for
the toroidal magnetic field evolution (background images, in {[}G{]}
units) for latitudes from 10$^{\circ}$ to 60$^{\circ}$, the radial
magnetic field is shown by contours in the range of $\pm10$G. The
right column shows the zonal acceleration (background images) and
the azimuthal velocity variations (shown by contours in the range
of $\pm3~$m/s.)}
\end{figure}

\begin{figure}
\includegraphics[width=0.75\textwidth]{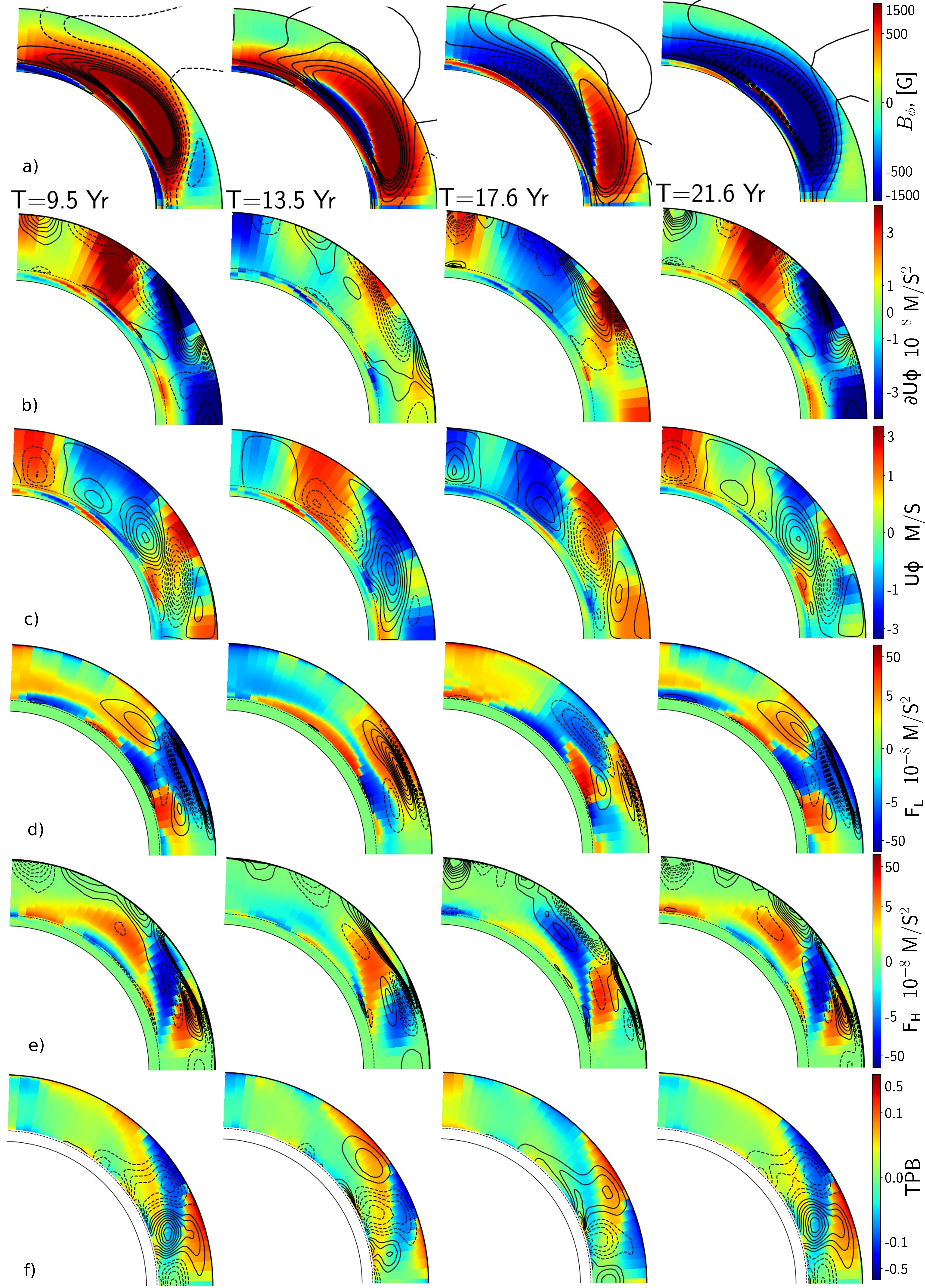} \caption{\label{czM1} Model M1. Snapshots for a half of dynamo cycle of: a)
the toroidal magnetic field (background image) and streamlines of
the poloidal field (contours); b) variations of the zonal acceleration
(background image) and the azimuthal force caused by variations of
the meridional circulation, $F_{U}$, (contour lines are plotted in
the range $\pm50$~m/s$^{2}$); c) variations of the azimuthal velosity
and contours show streamlines of the variations of meridional circulation,
dashed lines are for the counter clockwise circulation; d) the azimuthal
force caused by the Lorentz force $F_{L}^{(t)}$ (background image)
and the magnetic quenching of the turbulent angular momentum transport
$F_{\ell}$ (contours in the range of $\pm50$~m/s$^{2}$); e) the
dynamo-induced $\Lambda$-effect, $F_{H}$, 
 the term $H^{(0,\rho)}$ (background image), see Eqs~(\ref{h0}),
and contours (in the range $\pm50$~m/s$^{2}$) show variations of
the inertial forces, $F_{I}$; e) contours, in range of $\pm5\cdot10^{-3}$,
show the relative variations of the convective flux, $\delta F_{c}/F_{\odot}$
, where $F_{\odot}$ is the total energy flux, background image shows
the relative variations of the Taylor-Proudman balance (TPB); TPB
includes all terms of the right-hand side of Eq.~(\ref{eq:vort})
except the advection term. To display the whole range of the TPB variations
we use normalization to unity and the logarithmic scale}
\end{figure}

\begin{figure}
\includegraphics[width=0.45\textwidth]{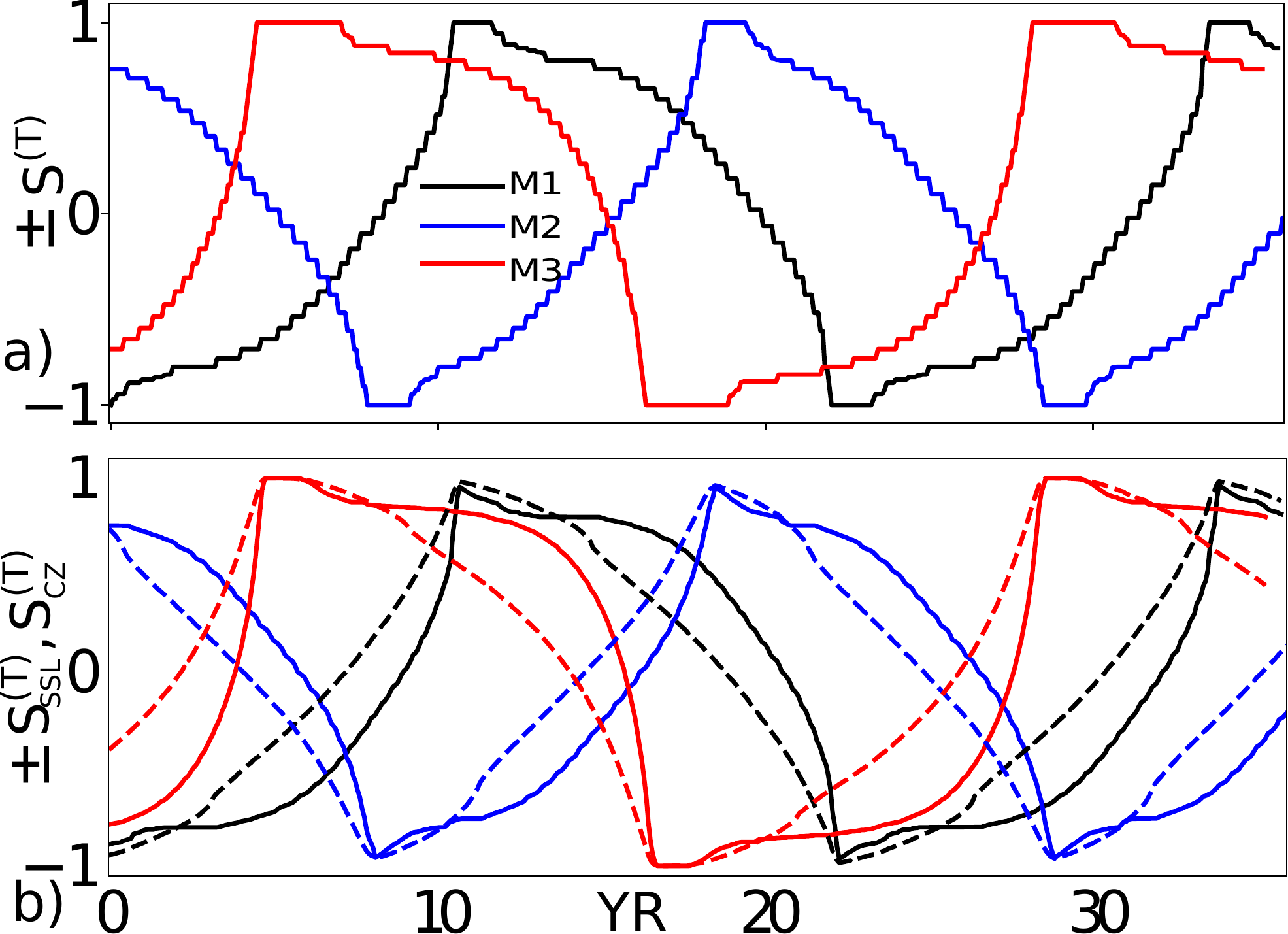} \caption{\label{fig:st}a) The relative area occupied by the toroidal magnetic
flux of one sign at the bottom of the subsurface shear layer, $\mathrm{S^{(T)}}$,
in models M1, M2 and M3; b) the same as (a) for the parameters $\mathrm{S_{SSL}^{(T)}}$(solid
lines) , and $\mathrm{S_{CZ}^{(T)}}$ (dashed lines).}
\end{figure}

\begin{figure}
\includegraphics[width=1\columnwidth]{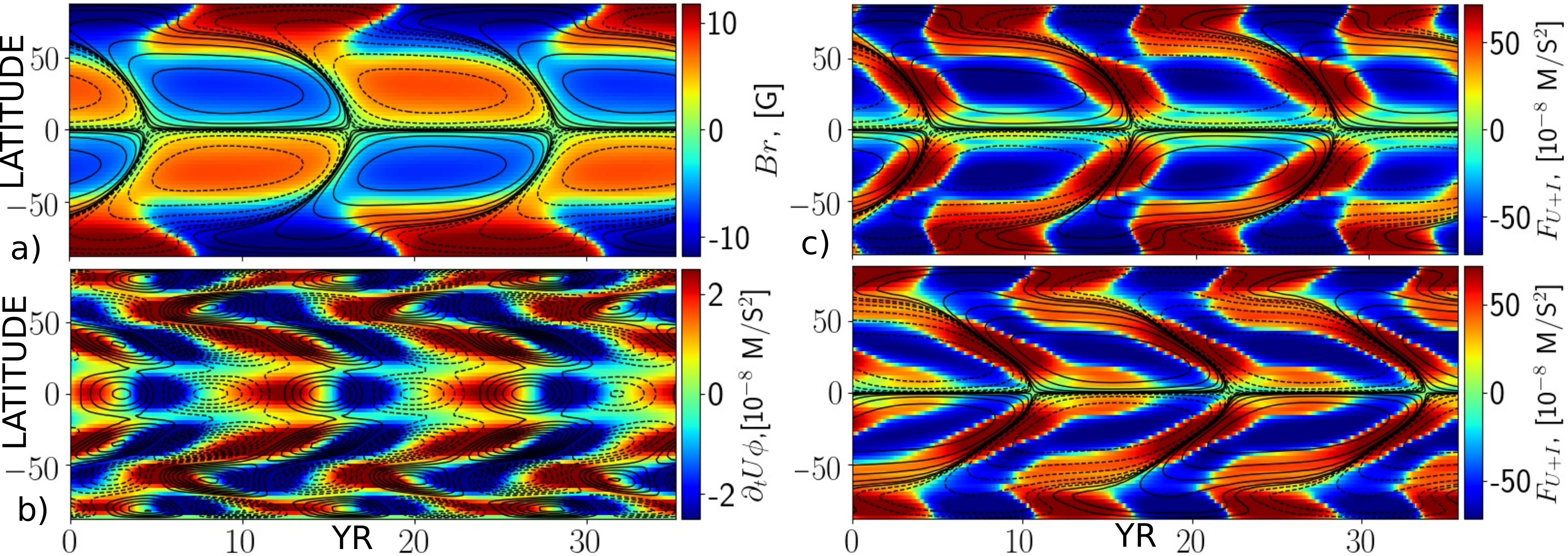} \caption{\label{fig:m13} Model M3. a) Time-latitude diagram of the radial
magnetic field at the surface (background image) and the toroidal
magnetic field at the bottom of the subsurface shear layer 
 b) the time-latitude diagram of the zonal flow acceleration (background
image) at the surface and the zonal velocity variations (contours
in range of $\pm3$m/s); c) the time-latitude diagram of $F_{I}+F_{U}$
(background image) for model M3; d) the same as c) for model M1, for
comparison.}
\end{figure}

\begin{figure}
\includegraphics[width=0.9\paperwidth]{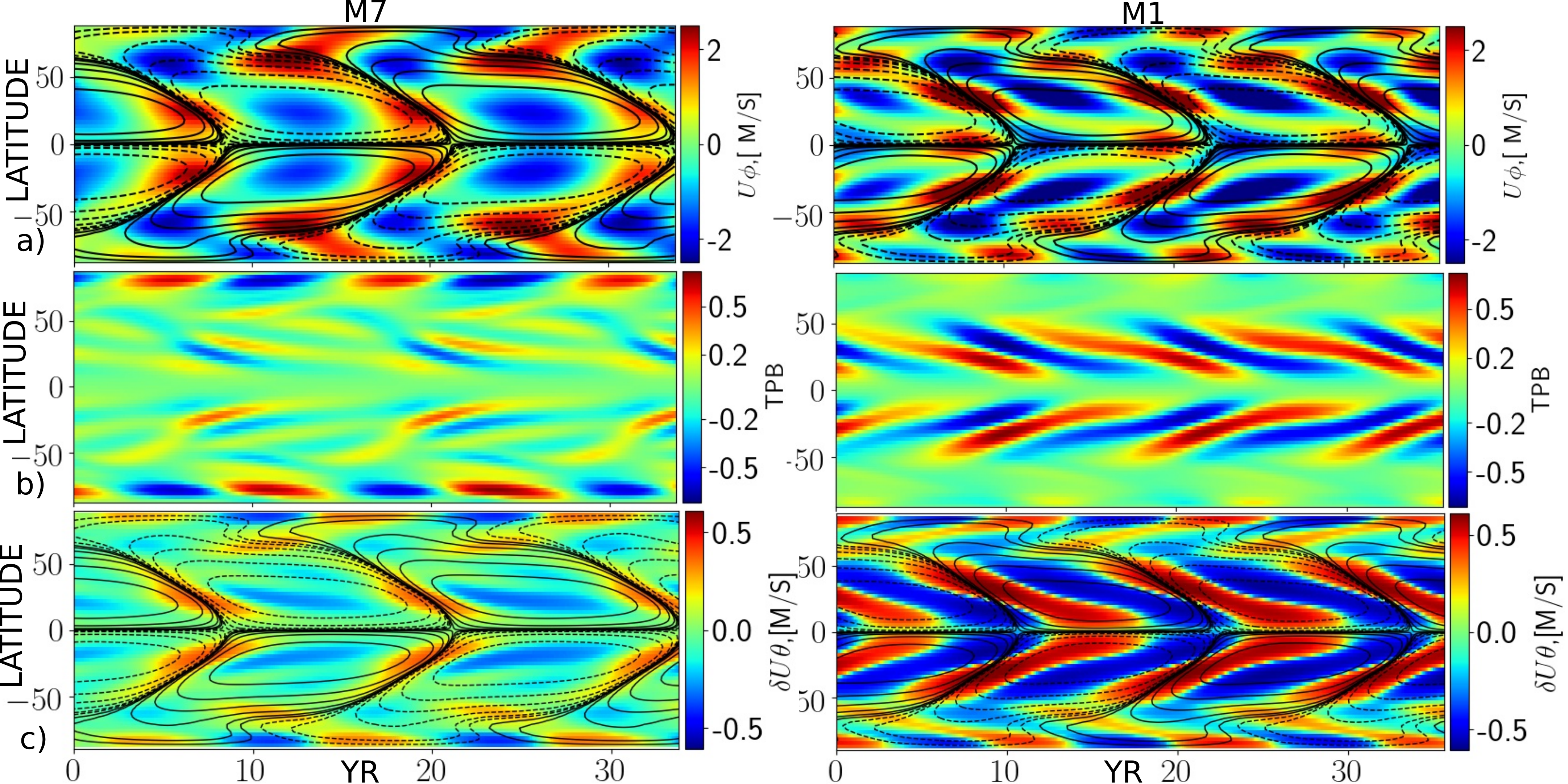} \caption{\label{fig:th} Model M7 (left column) and model M1 (right column).
a) Time-latitude diagram of the mean toroidal magnetic field in the
subsurface shear layer $r=0.9-0.99R$ (contour lines are plotted in
range of $\pm$1kG with \textit{exponential decrease of magnitude}
to the low values of $\pm$4G and the time-latitude diagram of the
azimuthal velocity oscillations (background image) at the surface;
b) the dynamo induced relative variations of the Taylor-Proudman balance
at the surface; c) the time-latitude diagram of the toroidal magnetic
field (same as in panel a) and variations of the meridional circulations
at the surface; positive variations mean the poleward flow. }
\end{figure}

\begin{figure}
\includegraphics[width=0.6\textwidth]{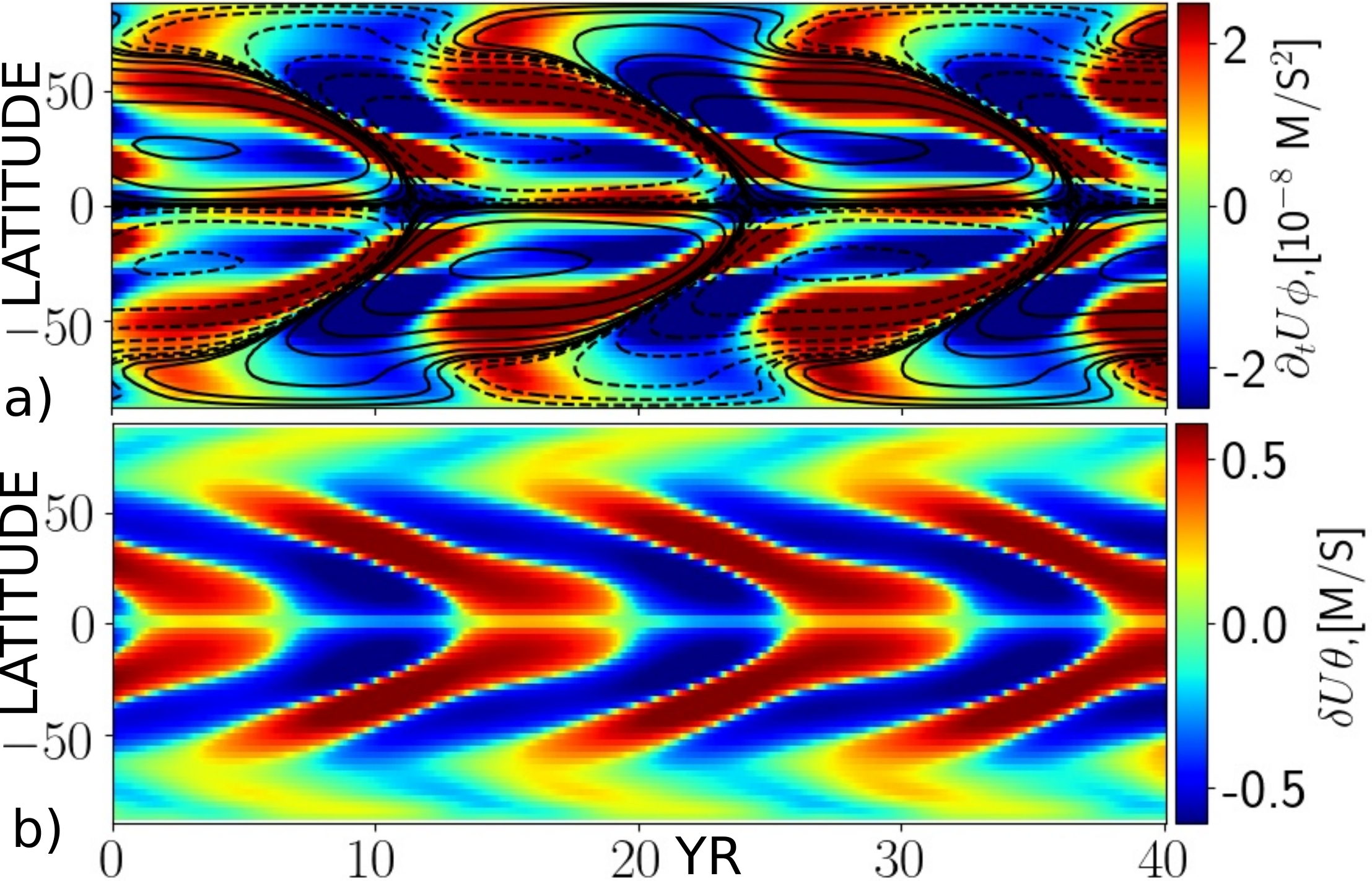} \caption{\textbf{\label{m8tl}}Model M8. a) Time-latitude diagram of the mean
toroidal magnetic field in the subsurface shear layer (contour lines
are plotted in range of $\pm$1kG with \textit{exponential decrease
of magnitude} to the low values of $\pm$4G) and the time-latitude
diagram of the azimuthal velocity acceleration (background image)
at the surface; b) variations of the meridional circulations at the
surface; positive variations mean the poleward flow.}
\end{figure}

\begin{figure}
\includegraphics[width=0.95\textwidth]{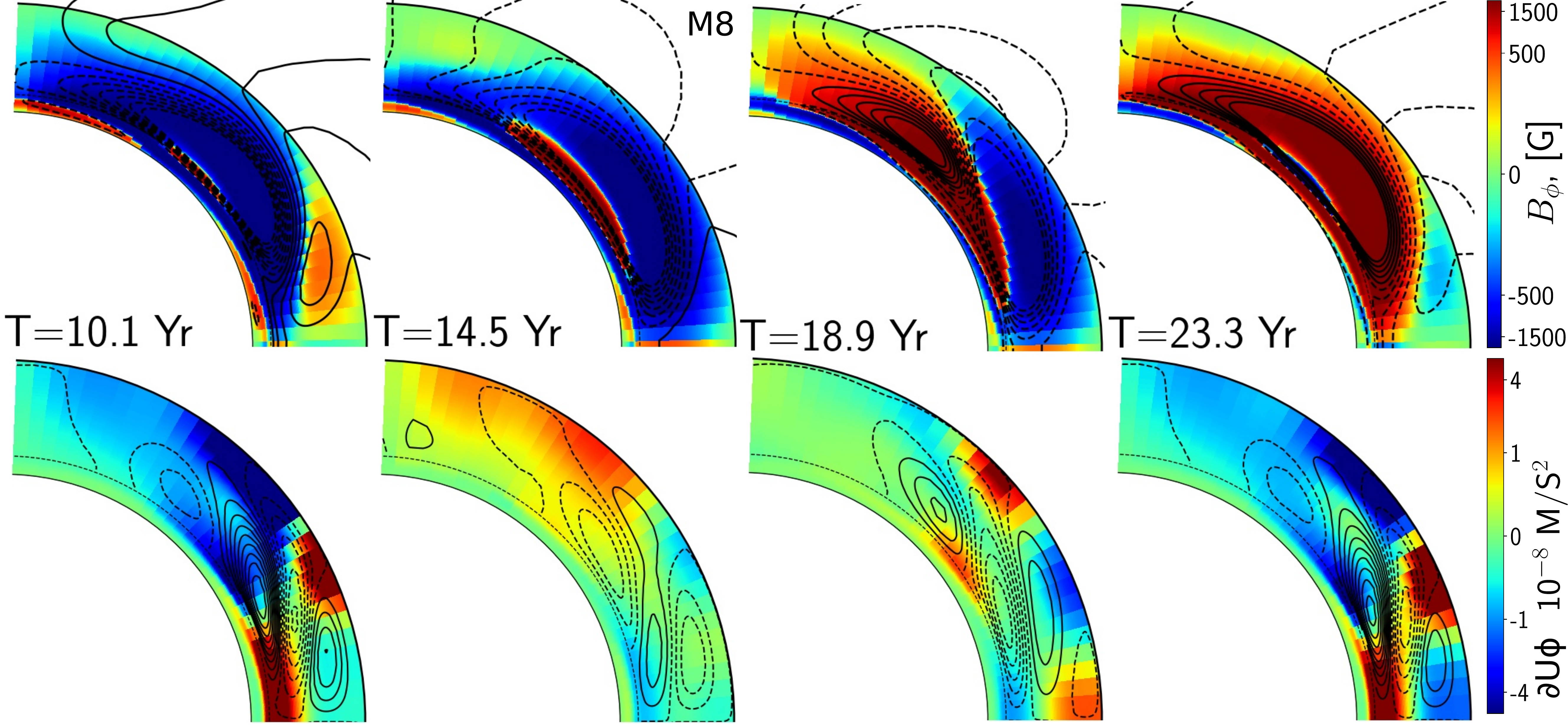} \caption{\label{m8sn} Model M8. a) Snapshots of the toroidal magnetic field
and streamlines of the poloidal magnetic field; b) snapshots of the
zonal accelerations (background images) and the streamlines of the
meridional circulation variations; amplitude of the meridional circulation
variation is about $\pm1$m/s.}
\end{figure}

The robustness of our conclusions depends on the radial distribution
of the large-scale field, which determines the zonal flow acceleration
in the convection zone. Comparing results of our model with observations
(Fig.~\ref{figure01}), we see that in our models the latitudinal
width of the zonal flow acceleration pattern is somewhat smaller than
in the observations . The models show axially aligned zonal acceleration
patterns at mid latitudes (Fig.~\ref{czM1}). The observations show
a slightly different inclination of these patterns \citep[e.g.,  Figure 2 in ][]{Kosovichev2019}.
The near-equatorial structure of the torsional oscillations is in
agreement with the observations, as well as variations of the meridional
circulation.

Summing up, it is found the extended 22-yr mode of the solar torsional
oscillations (zonal flow variations) can be explained by effects of
the overlapped dynamo waves on the angular momentum and heat transport
in the solar convection zone. The possible scenario of the extended
cycle of solar torsional oscillations is as follows. At high latitudes,
the torsional wave is excited  by the large-scale Lorentz force with
contributions from the dynamo-induced $\Lambda$-effect and the magnetic
quenching of the convective heat flux.
Then, the wave propagates towards the equator. Near the equator, the
large-scale Lorentz force and the dynamo-induced $\Lambda$-effect
are about zero. However, the toroidal magnetic field in the near-equatorial
(sunspot formation) zone results in quenching of the convective heat
flux. This induces variations of meridional flows from the polar and
equatorial sides to the central part of the sunspot formation zone.
These flows slow down the equatorial side of the dynamo wave, moving
the torsional wave closer to the equator. This meridional circulation
effect operates during the active phase of the cycle. Taking into
account the overlap of the dynamo waves, we explain the extended pattern
of the torsional waves.

\acknowledgements{ }

\indent Valery Pipin conducted the study as a part of FR II.16 of
ISTP SB RAS, the support of the RFBR-GFEN-a grant 19-52-53045 is acknowledged, as well.
Alexander Kosovichev was supported by the NASA grants: NNX14AB7CG and NNX17AE76A.

\bibliographystyle{aasjournal}

\begin{thebibliography}{}
\expandafter\ifx\csname natexlab\endcsname\relax\def\natexlab#1{#1}\fi
\providecommand{\url}[1]{\href{#1}{#1}}
\providecommand{\dodoi}[1]{doi:~\href{http://doi.org/#1}{\nolinkurl{#1}}}
\providecommand{\doeprint}[1]{\href{http://ascl.net/#1}{\nolinkurl{http://ascl.net/#1}}}
\providecommand{\doarXiv}[1]{\href{https://arxiv.org/abs/#1}{\nolinkurl{https://arxiv.org/abs/#1}}}

\bibitem[{{Altrock} {et~al.}(2008){Altrock}, {Howe}, \& {Ulrich}}]{Altrock2008}
{Altrock}, R., {Howe}, R., \& {Ulrich}, R. 2008, in Astronomical Society of the
  Pacific Conference Series, Vol. 383, Subsurface and Atmospheric Influences on
  Solar Activity, ed. R.~{Howe}, R.~W. {Komm}, K.~S. {Balasubramaniam}, \&
  G.~J.~D. {Petrie}, 335

\bibitem[{{Altrock}(1997)}]{Altrock1997}
{Altrock}, R.~C. 1997, \solphys, 170, 411, \dodoi{10.1023/A:1004958900477}

\bibitem[{{Beaudoin} {et~al.}(2013){Beaudoin}, {Charbonneau}, {Racine}, \&
  {Smolarkiewicz}}]{Beaudoin2013}
{Beaudoin}, P., {Charbonneau}, P., {Racine}, E., \& {Smolarkiewicz}, P.~K.
  2013, \solphys, 282, 335, \dodoi{10.1007/s11207-012-0150-2}

\bibitem[{{Bekki} \& {Yokoyama}(2017)}]{Bekki2017}
{Bekki}, Y., \& {Yokoyama}, T. 2017, \apj, 835, 9,
  \dodoi{10.3847/1538-4357/835/1/9}

\bibitem[{{B{\"o}ning} {et~al.}(2017){B{\"o}ning}, {Roth}, {Jackiewicz}, \&
  {Kholikov}}]{Boning2017}
{B{\"o}ning}, V.~G.~A., {Roth}, M., {Jackiewicz}, J., \& {Kholikov}, S. 2017,
  \apj, 845, 2, \dodoi{10.3847/1538-4357/aa7af0}

\bibitem[{Brandenburg(2005)}]{Brandenburg2005}
Brandenburg, A. 2005, Astrophys. J., 625, 539

\bibitem[{Brandenburg {et~al.}(1991)Brandenburg, Moss, R{\"{u}}diger, \&
  Tuominen}]{Brandenburg1991}
Brandenburg, A., Moss, D., R{\"{u}}diger, G., \& Tuominen, I. 1991, Geophysical
  {\&} Astrophysical Fluid Dynamics, 61, 179, \dodoi{10.1080/03091929108229043}

\bibitem[{{Brandenburg} {et~al.}(1992){Brandenburg}, {Moss}, \&
  {Tuominen}}]{Brandenburg1992}
{Brandenburg}, A., {Moss}, D., \& {Tuominen}, I. 1992, \aap, 265, 328

\bibitem[{{Cameron} \& {Sch{\"u}ssler}(2017)}]{Cameron17}
{Cameron}, R.~H., \& {Sch{\"u}ssler}, M. 2017, \aap, 599, A52,
  \dodoi{10.1051/0004-6361/201629746}

\bibitem[{{Choudhuri} \& {Dikpati}(1999)}]{Choudhuri1999}
{Choudhuri}, A.~R., \& {Dikpati}, M. 1999, \solphys, 184, 61

\bibitem[{{Covas} {et~al.}(2000){Covas}, {Tavakol}, {Moss}, \&
  {Tworkowski}}]{Covas2000}
{Covas}, E., {Tavakol}, R., {Moss}, D., \& {Tworkowski}, A. 2000, \aap, 360,
  L21

\bibitem[{{Dikpati} \& {Charbonneau}(1999)}]{Dikpati1999}
{Dikpati}, M., \& {Charbonneau}, P. 1999, \apj, 518, 508,
  \dodoi{10.1086/307269}

\bibitem[{{Durney}(1999)}]{Durney1999}
{Durney}, B.~R. 1999, \apj, 511, 945, \dodoi{10.1086/306696}

\bibitem[{{Guerrero} {et~al.}(2016{\natexlab{a}}){Guerrero}, {Smolarkiewicz},
  {de Gouveia Dal Pino}, {Kosovichev}, \& {Mansour}}]{Guerrero2016a}
{Guerrero}, G., {Smolarkiewicz}, P.~K., {de Gouveia Dal Pino}, E.~M.,
  {Kosovichev}, A.~G., \& {Mansour}, N.~N. 2016{\natexlab{a}}, \apjl, 828, L3,
  \dodoi{10.3847/2041-8205/828/1/L3}

\bibitem[{{Guerrero} {et~al.}(2016{\natexlab{b}}){Guerrero}, {Smolarkiewicz},
  {de Gouveia Dal Pino}, {Kosovichev}, \& {Mansour}}]{Guerrero2016}
---. 2016{\natexlab{b}}, \apj, 819, 104, \dodoi{10.3847/0004-637X/819/2/104}

\bibitem[{{Howe} {et~al.}(2018){Howe}, {Hill}, {Komm}, {Chaplin}, {Elsworth},
  {Davies}, {Schou}, \& {Thompson}}]{Howe2018}
{Howe}, R., {Hill}, F., {Komm}, R., {et~al.} 2018, \apj, 862, L5,
  \dodoi{10.3847/2041-8213/aad1ed}

\bibitem[{{Howe} {et~al.}(2011){Howe}, {Hill}, {Komm}, {Christensen-Dalsgaard},
  {Larson}, {Schou}, {Thompson}, \& {Ulrich}}]{Howe2011}
---. 2011, Journal of Physics Conference Series, 271, 012074,
  \dodoi{10.1088/1742-6596/271/1/012074}

\bibitem[{{Jouve} {et~al.}(2008){Jouve}, {Brun}, {Arlt}, {Brandenburg},
  {Dikpati}, {Bonanno}, {K{\"a}pyl{\"a}}, {Moss}, {Rempel}, {Gilman}, {Korpi},
  \& {Kosovichev}}]{Jouve2008}
{Jouve}, L., {Brun}, A.~S., {Arlt}, R., {et~al.} 2008, \aap, 483, 949,
  \dodoi{10.1051/0004-6361:20078351}

\bibitem[{{K{\"a}pyl{\"a}} {et~al.}(2016){K{\"a}pyl{\"a}}, {K{\"a}pyl{\"a}},
  {Olspert}, {Brandenburg}, {Warnecke}, {Karak}, \& {Pelt}}]{Kapyla2016}
{K{\"a}pyl{\"a}}, M.~J., {K{\"a}pyl{\"a}}, P.~J., {Olspert}, N., {et~al.} 2016,
  \aap, 589, A56, \dodoi{10.1051/0004-6361/201527002}

\bibitem[{{K{\"a}pyl{\"a}}(2019)}]{Kapyla2019}
{K{\"a}pyl{\"a}}, P.~J. 2019, \aap, 622, A195,
  \dodoi{10.1051/0004-6361/201732519}

\bibitem[{{Kichatinov}(1988)}]{Kichatinov1988}
{Kichatinov}, L.~L. 1988, Issledovaniia Geomagnetizmu Aeronomii i Fizike
  Solntsa, 82, 127

\bibitem[{{Kitchatinov} \& {Nepomnyashchikh}(2017)}]{Kitchatinov2017}
{Kitchatinov}, L.~L., \& {Nepomnyashchikh}, A.~A. 2017, Astronomy Letters, 43,
  332, \dodoi{10.1134/S106377371704003X}

\bibitem[{{Kitchatinov} \& {Olemskoy}(2011)}]{Kitchatinov2011a}
{Kitchatinov}, L.~L., \& {Olemskoy}, S.~V. 2011, Astronomy Letters, 37, 286,
  \dodoi{10.1134/S1063773711040037}

\bibitem[{{Kitchatinov} {et~al.}(1994){Kitchatinov}, {Pipin}, \&
  {Ruediger}}]{Kitchatinov1994}
{Kitchatinov}, L.~L., {Pipin}, V.~V., \& {Ruediger}, G. 1994, Astronomische
  Nachrichten, 315, 157

\bibitem[{Kitchatinov {et~al.}(1994)Kitchatinov, R\"udiger, \&
  Kueker}]{Kitchatinov1994a}
Kitchatinov, L.~L., R\"udiger, G., \& Kueker, M. 1994, \aap, 292, 125

\bibitem[{{Komm} {et~al.}(2012){Komm}, {Gonz{\'a}lez Hern{\'a}ndez}, {Hill},
  {Bogart}, {Rabello-Soares}, \& {Haber}}]{Komm2012}
{Komm}, R., {Gonz{\'a}lez Hern{\'a}ndez}, I., {Hill}, F., {et~al.} 2012,
  \solphys, 177, \dodoi{10.1007/s11207-012-0073-y}

\bibitem[{{Kosovichev} \& {Pipin}(2019)}]{Kosovichev2019}
{Kosovichev}, A.~G., \& {Pipin}, V.~V. 2019, \apj, 871, L20,
  \dodoi{10.3847/2041-8213/aafe82}

\bibitem[{{Kosovichev} {et~al.}(2013){Kosovichev}, {Pipin}, \&
  {Zhao}}]{Kosovichev2013}
{Kosovichev}, A.~G., {Pipin}, V.~V., \& {Zhao}, J. 2013, in Astronomical
  Society of the Pacific Conference Series, Vol. 479, Progress in Physics of
  the Sun and Stars: A New Era in Helio- and Asteroseismology, ed.
  H.~{Shibahashi} \& A.~E. {Lynas-Gray}, 395

\bibitem[{{Kosovichev} \& {Zhao}(2016)}]{Kosovichev2016}
{Kosovichev}, A.~G., \& {Zhao}, J. 2016, in Lecture Notes in Physics, Berlin
  Springer Verlag, Vol. 914, Lecture Notes in Physics, Berlin Springer Verlag,
  ed. J.-P. {Rozelot} \& C.~{Neiner}, 25

\bibitem[{{Kosovichev} {et~al.}(1997){Kosovichev}, {Schou}, {Scherrer},
  {Bogart}, {Bush}, {Hoeksema}, {Aloise}, {Bacon}, {Burnette}, {de Forest},
  {Giles}, {Leibrand}, {Nigam}, {Rubin}, {Scott}, {Williams}, {Basu},
  {Christensen-Dalsgaard}, {Dappen}, {Rhodes}, {Duvall}, {Howe}, {Thompson},
  {Gough}, {Sekii}, {Toomre}, {Tarbell}, {Title}, {Mathur}, {Morrison}, {Saba},
  {Wolfson}, {Zayer}, \& {Milford}}]{Kosovichev1997}
{Kosovichev}, A.~G., {Schou}, J., {Scherrer}, P.~H., {et~al.} 1997, \solphys,
  170, 43, \dodoi{10.1023/A:1004949311268}

\bibitem[{Krause \& R\"adler(1980)}]{Krause1980}
Krause, F., \& R\"adler, K.-H. 1980, Mean-Field Magnetohydrodynamics and Dynamo
  Theory (Berlin: Akademie-Verlag), 271

\bibitem[{{Kueker} {et~al.}(1996){Kueker}, {Ruediger}, \& {Pipin}}]{Kueker1996}
{Kueker}, M., {Ruediger}, G., \& {Pipin}, V.~V. 1996, \aap, 312, 615

\bibitem[{{K{\"u}ker} {et~al.}(1999){K{\"u}ker}, {Arlt}, \&
  {R{\"u}diger}}]{Kuker1999}
{K{\"u}ker}, M., {Arlt}, R., \& {R{\"u}diger}, G. 1999, \aap, 343, 977

\bibitem[{{Labonte} \& {Howard}(1982)}]{Labonte1982}
{Labonte}, B.~J., \& {Howard}, R. 1982, \solphys, 75, 161,
  \dodoi{10.1007/BF00153469}

\bibitem[{{Ludwig} {et~al.}(1999){Ludwig}, {Freytag}, \&
  {Steffen}}]{Ludwig1999}
{Ludwig}, H.-G., {Freytag}, B., \& {Steffen}, M. 1999, \aap, 346, 111

\bibitem[{{Malkus} \& {Proctor}(1975)}]{Malkus1975}
{Malkus}, W.~V.~R., \& {Proctor}, M.~R.~E. 1975, Journal of Fluid Mechanics,
  67, 417, \dodoi{10.1017/S0022112075000390}

\bibitem[{{Miesch} {et~al.}(2011){Miesch}, {Brown}, {Browning}, {Brun}, \&
  {Toomre}}]{Miesch2011a}
{Miesch}, M.~S., {Brown}, B.~P., {Browning}, M.~K., {Brun}, A.~S., \& {Toomre},
  J. 2011, in IAU Symposium, Vol. 271, IAU Symposium, ed. N.~H. {Brummell},
  A.~S. {Brun}, M.~S. {Miesch}, \& Y.~{Ponty}, 261--269

\bibitem[{{Moss} \& {Brandenburg}(1992)}]{Moss1992}
{Moss}, D., \& {Brandenburg}, A. 1992, \aap, 256, 371

\bibitem[{Parker(1955)}]{Parker1955}
Parker, E. 1955, Astrophys. J., 122, 293

\bibitem[{{Paxton} {et~al.}(2011){Paxton}, {Bildsten}, {Dotter}, {Herwig},
  {Lesaffre}, \& {Timmes}}]{Paxton2011}
{Paxton}, B., {Bildsten}, L., {Dotter}, A., {et~al.} 2011, \apjs, 192, 3,
  \dodoi{10.1088/0067-0049/192/1/3}

\bibitem[{{Paxton} {et~al.}(2013){Paxton}, {Cantiello}, {Arras}, {Bildsten},
  {Brown}, {Dotter}, {Mankovich}, {Montgomery}, {Stello}, {Timmes}, \&
  {Townsend}}]{Paxton2013}
{Paxton}, B., {Cantiello}, M., {Arras}, P., {et~al.} 2013, \apjs, 208, 4,
  \dodoi{10.1088/0067-0049/208/1/4}

\bibitem[{{Pipin}(1999)}]{Pipin1999}
{Pipin}, V.~V. 1999, \aap, 346, 295

\bibitem[{{Pipin}(2003)}]{Pipin2003}
---. 2003, Geophysical and Astrophysical Fluid Dynamics, 97, 25,
  \dodoi{10.1080/0309192021000053366}

\bibitem[{{Pipin}(2004)}]{Pipin2004}
---. 2004, Astronomy Reports, 48, 418, \dodoi{10.1134/1.1744942}

\bibitem[{{Pipin}(2008)}]{Pipin2008a}
---. 2008, Geophysical and Astrophysical Fluid Dynamics, 102, 21

\bibitem[{{Pipin}(2018)}]{Pipin2018b}
---. 2018, Journal of Atmospheric and Solar-Terrestrial Physics, 179, 185,
  \dodoi{10.1016/j.jastp.2018.07.010}

\bibitem[{{Pipin} \& {Kitchatinov}(2000)}]{Pipin2000}
{Pipin}, V.~V., \& {Kitchatinov}, L.~L. 2000, Astronomy Reports, 44, 771,
  \dodoi{10.1134/1.1320504}

\bibitem[{{Pipin} \& {Kosovichev}(2011{\natexlab{a}})}]{Pipin2011a}
{Pipin}, V.~V., \& {Kosovichev}, A.~G. 2011{\natexlab{a}}, ApJL, 727, L45,
  \dodoi{10.1088/2041-8205/727/2/L45}

\bibitem[{{Pipin} \& {Kosovichev}(2011{\natexlab{b}})}]{Pipin2011}
---. 2011{\natexlab{b}}, ApJ, 741, 1, \dodoi{10.1088/0004-637X/741/1/1}

\bibitem[{{Pipin} \& {Kosovichev}(2014)}]{Pipin2014}
---. 2014, \apj, 785, 49, \dodoi{10.1088/0004-637X/785/1/49}

\bibitem[{{Pipin} \& {Kosovichev}(2018)}]{Pipin2018c}
---. 2018, \apj, 854, 67, \dodoi{10.3847/1538-4357/aaa759}

\bibitem[{{Pipin} {et~al.}(2012){Pipin}, {Sokoloff}, \& {Usoskin}}]{Pipin2012}
{Pipin}, V.~V., {Sokoloff}, D.~D., \& {Usoskin}, I.~G. 2012, \aap, 542, A26,
  \dodoi{10.1051/0004-6361/201118733}

\bibitem[{{Rajaguru} \& {Antia}(2015)}]{Rajaguru2015}
{Rajaguru}, S.~P., \& {Antia}, H.~M. 2015, \apj, 813, 114,
  \dodoi{10.1088/0004-637X/813/2/114}

\bibitem[{{Rempel}(2006)}]{Rempel2006}
{Rempel}, M. 2006, \apj, 647, 662, \dodoi{10.1086/505170}

\bibitem[{{Rempel}(2007)}]{Rempel2007ApJ}
---. 2007, \apj, 655, 651, \dodoi{10.1086/509866}

\bibitem[{{Rogachevskii} \& {Kleeorin}(2018)}]{Rogachevskii2018}
{Rogachevskii}, I., \& {Kleeorin}, N. 2018, Journal of Plasma Physics, 84,
  735840201, \dodoi{10.1017/S0022377818000272}

\bibitem[{{Ruediger}(1989)}]{Ruediger1989}
{Ruediger}, G. 1989, Differential rotation and stellar convection. Sun and the
  solar stars (Akademie-Verlag, Berlin)

\bibitem[{{Ruediger} \& {Brandenburg}(1995)}]{Ruediger1995}
{Ruediger}, G., \& {Brandenburg}, A. 1995, \aap, 296, 557

\bibitem[{{Snodgrass} \& {Howard}(1985)}]{Snodgrass1985}
{Snodgrass}, H.~B., \& {Howard}, R. 1985, \solphys, 95, 221,
  \dodoi{10.1007/BF00152399}

\bibitem[{{Spruit}(2003)}]{Spruit2003}
{Spruit}, H.~C. 2003, \solphys, 213, 1, \dodoi{10.1023/A:1023202605379}

\bibitem[{{Stenflo}(1992)}]{Sten1992}
{Stenflo}, J.~O. 1992, in Astronomical Society of the Pacific Conference
  Series, Vol.~27, The Solar Cycle, ed. K.~L. {Harvey}, 421

\bibitem[{{Stenflo} \& {Guedel}(1988)}]{Stenflo1988}
{Stenflo}, J.~O., \& {Guedel}, M. 1988, \aap, 191, 137

\bibitem[{{Stix}(2002)}]{Stix2002}
{Stix}, M. 2002, The sun: an introduction, 2nd edn. (Berlin : Springer), 521

\bibitem[{{Ulrich}(2001)}]{Ulrich2001}
{Ulrich}, R.~K. 2001, \apj, 560, 466, \dodoi{10.1086/322524}

\bibitem[{{Wilson} {et~al.}(1988){Wilson}, {Altrocki}, {Harvey}, {Martin}, \&
  {Snodgrass}}]{Wilson1988}
{Wilson}, P.~R., {Altrocki}, R.~C., {Harvey}, K.~L., {Martin}, S.~F., \&
  {Snodgrass}, H.~B. 1988, \nat, 333, 748, \dodoi{10.1038/333748a0}

\bibitem[{{Wright} \& {Drake}(2016)}]{Wright2016}
{Wright}, N.~J., \& {Drake}, J.~J. 2016, \nat, 535, 526,
  \dodoi{10.1038/nature18638}

\bibitem[{{Yoshimura}(1981)}]{Yoshimura1981}
{Yoshimura}, H. 1981, \apj, 247, 1102, \dodoi{10.1086/159120}

\bibitem[{{Zhao} {et~al.}(2014){Zhao}, {Kosovichev}, \& {Bogart}}]{Zhao2014}
{Zhao}, J., {Kosovichev}, A.~G., \& {Bogart}, R.~S. 2014, \apjl, 789, L7,
  \dodoi{10.1088/2041-8205/789/1/L7}

\end{thebibliography}

 \clearpage{}

\begin{table}
\caption{\label{tab:Models}Model parameters: $\eta_{A}$ controls the anisotropic
eddy diffusivity (see, Eq.~\ref{eq:diff}); $\ell_{\mathrm{min}}$
controls the number of meridional circulation cells along the radius;
columns $F_{U}$, $F_{\ell}$, $F_{L}^{(\mathrm{t,p})}$, and $F_{H}$
show the nonlinear and dynamic effects involved in the angular momentum
balance; contribution $H^{(0,a)}$ stems from anisotropy of convective
turbulence, and $H^{(0,\rho)}$ results from the density stratification
(see the text).}

\begin{tabular}{ccccccccc}
\hline 
Model  & $\eta_{A}$  & $\mathrm{\ell_{min}}/R$  & N Cells  & $F_{U}$  & $F_{\ell}$  & $F_{L}^{(\mathrm{t,p})}$  & $\mathrm{\chi_{ij}}$(Eqs(\ref{conv},\ref{eq:htfl}))  & $F_{H}$\tabularnewline
\hline 
M1  & 0  & $0.02$  & 1  & Y  & Y  & Y  & $\mathrm{\chi_{ij}\left(\Omega^{\star},\beta\right)}$  & $H^{(0,\rho)}$ \tabularnewline
\hline 
M2  & 0  & 0.01  & 2  & -/-  & -/-  & -/-  & -/-  & $H^{(0,\rho)}$ \tabularnewline
\hline 
M3  & 2$\eta_{T}$  & 0.02  & 1  & -/-  & -/-  & -/-  & -/-  & $H^{(0,\rho)}$ \tabularnewline
\hline 
M4  & 0  & 0.02  & 1  & Y  & N  & Y  & -/-  & 0 \tabularnewline
\hline 
M5  & 0  & -/-  & 1  & Y  & Y  & N  & -/-  & $H^{(0,\rho)}$ \tabularnewline
\hline 
M6  & 0  & -/-  & 1  & Y  & Y  & N  & $\mathrm{\chi_{ij}\left(\Omega^{\star},\beta\right)}$  & $H^{(0,a)}$\tabularnewline
\hline 
M7  & 0  & -/-  & 1  & Y  & Y  & Y  & $\mathrm{\chi_{ij}\left(\Omega^{\star}\right)}$  & $H^{(0,\rho)}$\tabularnewline
\hline 
M8  & 0  & 0.02  & 1  & Y  & N  & N  & $\mathrm{\chi_{ij}\left(\Omega^{\star},\beta\right)}$  & 0\tabularnewline
\hline 
\end{tabular}
\end{table}

\begin{table}
\caption{\label{tab:Models-r}Model properties: the dynamo cycle period, the
delay between the subsequent cycles (see the main text), duration
of the extended dynamo mode, magnitudes of the torsional oscillation
velocity and zonal acceleration on the surface, duration of the equatorial
branch of the extended mode of the torsional oscillations, duration
of the polar branch of the torsional oscillations.}

\begin{tabular}{>{\centering}p{1cm}>{\centering}p{1.5cm}>{\centering}p{1cm}>{\centering}p{1.5cm}>{\centering}p{1cm}>{\centering}p{1.5cm}>{\centering}p{2cm}>{\centering}p{2cm}}
\hline 
Observ./ Model  & Cycle Period, {[}yr{]}  & Cycle's $\Delta t$, {[}yr{]}  & Dynamo Ext. mode Length, {[}yr{]}  & $\delta U_{\phi}$ {[}M/S{]}  & $\partial_{t}U_{\phi}$ $10^{-8}${[}M/S$^{2}${]}  & Ext. mode, {[}yr{]}  & Polar branch, {[}yr{]}\tabularnewline
\hline 
Observ.  & 22  &  & 20  &  & $\pm8$  & 20  & 9\tabularnewline
\hline 
M1  & 22.9  & 1.2  & 21.7  & $\pm4.3$  & $\pm5.6$  & 24.  & 8\tabularnewline
\hline 
M2  & 20.6  & 1.1  & 19.2  & $\pm4.7$  & $\pm7.1$  & 22  & 7.8\tabularnewline
\hline 
M3  & 23.4  & 2.6  & 23.5  & $\pm2.7$  & $\pm4.5$  & N  & 6\tabularnewline
\hline 
M4  & 24.5  & 1.2  & 25.1  & $\pm3$  & $\pm6.8$  & 25.1  & 10\tabularnewline
\hline 
M5  & 22.6  & 0.9  & 23.8  & $\pm4$  & $\pm7$  & 20  & 8.1\tabularnewline
\hline 
M6  & 20.1  & 0.9  & 20.1  & $\pm5.7$  & $\pm14.2$  & N  & 5\tabularnewline
\hline 
M7  & 25  & 1.8  & 26.3  & $\pm2.7$  & $\pm4.4$  & N  & 9\tabularnewline
\hline 
M8  & 25.1  & 1.7  & 25  & $\pm4$  & $\pm7$  & 23.1  & 8.3\tabularnewline
\hline 
\end{tabular}
\end{table}

\clearpage{} 

\clearpage{}

\appendix

\section{Model Details}

A detailed description of the mean-field solar dynamo model, as well
as the model for the angular momentum balance and heat transport,
can be found in our previous papers \citep{Pipin2018b,Pipin2018c}.
Here, we collect analytical results on the magnetic field contribution
to the $\Lambda$-effect. This contribution is a part of the turbulent
stress tensor 
\begin{equation}
{\hat{T}_{ij}=\left(\left\langle u_{i}u_{j}\right\rangle -\frac{1}{4\pi\overline{\rho}}\left(\left\langle b_{i}b_{j}\right\rangle -\frac{1}{2}\delta_{ij}\left\langle \mathbf{b}^{2}\right\rangle \right)\right),}\label{eq:stres-1}
\end{equation}
where $\mathbf{u}$ and $\mathbf{b}$ are fluctuating velocity and
magnetic fields. Application the mean-field hydrodynamic framework
(see, \citealt{Kitchatinov1994}) leads to the Taylor expansion in
terms of the scale-separation parameter $\ell/L$: 
\begin{eqnarray}
\hat{T}_{ij} & = & \hat{T}_{ij}^{\left(\Lambda\right)}+\hat{T}_{ij}^{\left(\nu\right)}\label{eq:tstr}\\
 & = & \Lambda_{ijk}\Omega_{k}-N_{ijkl}\frac{\partial\overline{U}_{k}}{\partial r_{l}}+\dots
\end{eqnarray}
The non-diffusive flux of angular momentum $\boldsymbol{\Lambda}=\left\langle {u}_{\phi}'{\boldsymbol{u}}\right\rangle $
can be expressed as follows \citep{Ruediger1989}: 
\begin{eqnarray}
\hat{T}_{r\phi}^{\left(\Lambda\right)} & = & \Lambda_{V}\Omega\sin\theta,\nonumber \\
\Lambda_{V} & = & \nu_{T}\left(V^{(0)}+\sin^{2}\theta V^{(1)}\right),\label{eq:lv}\\
\hat{T}_{\theta\phi}^{\left(\Lambda\right)} & = & \Lambda_{H}\Omega\cos\theta,\nonumber \\
\Lambda_{H} & = & \nu_{T}\left(H^{(0)}+\sin^{2}\theta H^{(1)}\right)\label{eq:lh}
\end{eqnarray}
In the previous papers, we discuss in detail the standard parts of
this parametrization for coefficients $V^{(0)}$, $V^{(1)}$ and $H^{(1)}$.
It was shown that the analytical form of the $\Lambda$-effect coefficients
becomes fairly complicated if we wish to account for the multiple-cell
meridional circulation structure. Also, results of \citet{Pipin2018c}
show that the spatial derivative of the Coriolis number $\Omega^{*}=2\Omega_{0}\tau_{c}$,
has to be taken into account.

In our nonlinear models, we take into account the effect of magnetic
field (see, \citealt{Pipin1999}), as well as the effect of convective
velocities anisotropy, which is important for modeling the subsurface
shear layer. Therefore, the final coefficients of the $\Lambda$-tensor
are: 
\begin{eqnarray}
V^{(0)} & = & \Bigl[\left(\frac{\alpha_{MLT}}{\gamma}\right)^{2}\left\{ J_{0}+\!J_{1}+\!a\left(I_{0}+\!I_{1}\right)\right\} \label{v0-f}\\
 & - & \Bigl(\frac{\alpha_{MLT}\ell}{\gamma}\frac{\partial}{\partial r}\left\{ \left(J_{0}+J_{1}\right)-I_{5}+I_{6}\right\} +\ell^{2}\frac{\partial^{2}}{\partial r^{2}}\left(I_{1}-I_{2}\right)\Bigr]\phi_{\chi}^{(I)}\left(\beta\right),\nonumber \\
V^{(1)} & = & -\left\{ \left(\frac{\alpha_{MLT}}{\gamma}\right)^{2}\left(J_{1}+aI_{1}\right)-\frac{\alpha_{MLT}\ell}{\gamma}\frac{\partial}{\partial r}\left(J_{1}+I_{6}\right)-\ell^{2}\frac{\partial^{2}}{\partial r^{2}}I_{2}\right\} \phi_{\chi}^{(I)}\left(\beta\right),\label{v1-f}
\end{eqnarray}
and ${H^{(1)}=-V^{(1)}}$. We introduce the anisotropy parameter,
$a={\displaystyle \frac{\overline{u_{h}^{2}}-2\overline{u_{r}^{2}}}{\overline{u_{r}^{2}}}=}2$,
where $u_{h}$ and $u_{r}$ are the horizontal and vertical RMS velocities.
Collecting results of \citet{Kitchatinov1994a} and \citet{Kueker1996},
we write the coefficient, $H^{(0,\rho)}$, as follows: 
\begin{eqnarray}
H^{(0,\rho)} & = & \frac{\tau^{2}\left\langle \boldsymbol{u}'^{2}\right\rangle }{\rho^{2}}\phi_{H}\left(\beta\right)\frac{\partial^{2}}{\partial r^{2}}\left(\rho^{2}J_{4}\right)\label{h0-f}\\
 & = & \left\{ 4\left(\frac{\alpha_{MLT}}{\gamma}\right)^{2}J_{4}-4\frac{\alpha_{MLT}\ell}{\gamma}\frac{\partial}{\partial r}J_{4}+\ell^{2}\frac{\partial^{2}}{\partial r^{2}}J_{4}\right\} \phi_{H}\left(\beta\right),\nonumber 
\end{eqnarray}
where function $J_{4}$ was defined by \citet{Kitchatinov1994a},
and the magnetic quenching function $\phi_{H}\left(\beta\right)$
was defined by \citet{Pipin2003}: 
\begin{eqnarray}
\phi_{H} & = & \frac{1}{\beta^{2}}\left(\frac{2+3\beta^{2}}{2\sqrt{\left(1+\beta^{2}\right)^{3}}}-1\right).
\end{eqnarray}
The magnetically induced $\Lambda$-effect caused by the turbulence
anisotropy was discussed by \citet{Kichatinov1988}. Following his
results, we define: 
\begin{eqnarray}
H^{(0,a)} & = & \left(\frac{\alpha_{MLT}}{\gamma}\right)^{2}aJ_{H}\left(\beta\right)J_{3}\left(\Omega^{*}\right)\label{h0a}
\end{eqnarray}
where 
\begin{eqnarray}
J_{H} & = & \frac{105}{256\beta^{4}}\left(\beta^{2}-37-\frac{8}{1+\beta^{2}}+\left(\beta^{4}+6\beta^{2}+45\right)\frac{\arctan\beta}{\beta}\right),\\
J_{3} & = & \frac{9}{15\Omega^{*6}}\left(15+\Omega^{*2}+\left(\Omega^{*4}-6\Omega^{*2}-15\right)\frac{\arctan\Omega^{*}}{\Omega^{*}}\right).
\end{eqnarray}

The convective heat transport is treated in the same way as in the
recent paper of \citet{Pipin2018b}. For expressions of the eddy heat
conductivity tensor we employ the following expression: 
\begin{equation}
\chi_{ij}=\chi_{T}\left(\phi_{\chi}^{(I)}\left(\beta\right)\phi\left(\Omega^{*}\right)\delta_{ij}+\phi_{\chi}^{(\|)}\left(\beta\right)\phi_{\parallel}\left(\Omega^{*}\right)\frac{\Omega_{i}\Omega_{j}}{\Omega^{2}}\right),\label{eq:htfl}
\end{equation}
where functions $\phi$ and $\phi_{\parallel}$ were defined in \citet{Kitchatinov1994},
and the magnetic quenching functions $\mathrm{\phi_{\chi}^{(I)}}$
and $\phi_{\chi}^{(\|)}$ are 
\begin{eqnarray*}
\mathrm{\phi_{\chi}^{(I)}} & \mathrm{=} & \mathrm{\frac{2}{\beta^{2}}\left(1-\frac{1}{\sqrt{1+\beta^{2}}}\right),}\\
\phi_{\chi}^{(\|)} & = & \mathrm{\frac{2}{\beta^{2}}\left(\sqrt{1+\beta^{2}}-1\right).}
\end{eqnarray*}

\end{document}